\documentclass[journal]{IEEEtran}
\IEEEoverridecommandlockouts
\usepackage[space]{cite}
\usepackage{authblk}
\usepackage{amsmath, amsthm, amssymb, amsfonts, bm, mathtools}
\usepackage{caption}
\usepackage[skip = 0pt]{subcaption}
\usepackage{dblfloatfix}
\newtheorem{theorem}{Theorem}
\newtheorem{lemma}{Lemma}
\newtheorem{corollary}[theorem]{Corollary}
\newtheorem{proposition}[theorem]{Proposition}
\newtheorem{assumption}{Assumption}
\usepackage{algorithm}
\usepackage{algorithmic}

\newcommand{\norm}[1]{\lVert#1\rVert}
\newcommand{\s}{\hspace{.25cm}}

\usepackage{hyperref}
\hypersetup{
    colorlinks=true,
    linkcolor=blue,
    filecolor=magenta,      
    urlcolor=cyan
    }

\usepackage{fancyhdr}

\fancypagestyle{firstpagestyle}
{
\fancyhf{}
\rhead{\scriptsize \thepage}
\cfoot{\footnotesize This work has been submitted to the IEEE for possible publication. Copyright may be transferred without notice, after which this version may no longer be accessible.}

}

\begin{document}
\bstctlcite{IEEEexample:BSTcontrol}

\title{Linear Inverse Problems Using a Generative Compound Gaussian Prior
}
\author{{\large Carter Lyons, Raghu G. Raj, and Margaret Cheney}
\thanks{Carter Lyons is with the U.S. Naval Research Laboratory, Washington, DC 20375 USA and also with Colorado State University, Fort Collins, CO 80523 USA (e-mail: carter.lyons@colostate.edu)}
\thanks{Raghu G. Raj is with the U.S. Naval Research Laboratory, Washington, DC 20375 USA (e-mail: raghu.g.raj@ieee.org)}
\thanks{Margaret Cheney is with Colorado State University, Fort Collins, CO 80523 USA (e-mail: margaret.cheney@colostate.edu)}
\thanks{This material is based upon research supported in part by the U.S. Office of Naval Research via the NRL base program and under award numbers N000142412241 and N0001423WX01875, and sponsored in part by the Air Force Office of Scientific Research under award number FA9550-21-1-0169.}
}

\maketitle
\thispagestyle{firstpagestyle}

\begin{abstract}
Since most inverse problems arising in scientific and engineering applications are ill-posed, prior information about the solution space is incorporated, typically through regularization, to establish a well-posed problem with a unique solution. Often, this prior information is an assumed statistical distribution of the desired inverse problem solution. Recently, due to the unprecedented success of generative adversarial networks (GANs), the generative network from a GAN has been implemented as the prior information in imaging inverse problems. In this paper, we devise a novel iterative algorithm to solve inverse problems in imaging where a dual-structured prior is imposed by combining a GAN prior with the compound Gaussian (CG) class of distributions. A rigorous computational theory for the convergence of the proposed iterative algorithm, which is based upon the alternating direction method of multipliers, is established. Furthermore, elaborate empirical results for the proposed iterative algorithm are presented. By jointly exploiting the powerful CG and GAN classes of image priors, we find, in compressive sensing and tomographic imaging problems, our proposed algorithm outperforms and provides improved generalizability over competitive prior art approaches while avoiding performance saturation issues in previous GAN prior-based methods. 
\end{abstract}


\section{Introduction}

\subsection{Motivation}

\IEEEPARstart{I}{nverse} problems (IPs), the problem of recovering input model parameters -- e.g., an image --  representing a physical quantity from observed output data -- e.g., an undersampling or corruption -- produced by a forward measurement mapping, arise throughout many applications in scientific and engineering fields. In imaging, for example, compressive sensing (CS), X-ray computed tomography (CT), magnetic resonance imaging (MRI), super-resolution, radar imaging, and sonar imaging are just a handful of practical applications of linear IPs where the forward mapping, from input model parameters to output data, is effectively captured by a linear operator. Typically, forward mappings arising from scientific and engineering applications are non-injective, creating (linear) IPs that are ill-posed since the forward mapping sends multiple distinct input model parameterizations to the same observed output data. To alleviate ill-posedness, outside prior information about the input model parameters -- e.g., sparsity or known statistics -- is incorporated into the solution of the (linear) IPs. That is, given observed output data, the solution to the (linear) IP is input model parameters that, in addition to mapping to the observed output data via the forward measurement mapping, satisfies statistical or structural conditions imposed by the prior information.

Model-based methods -- in particular, within a Bayesian maximum a posteriori (MAP) framework -- are a primary approach of solving (linear) IPs and often involve the minimization of a cost function via an iterative algorithm. Prior information is encapsulated by the cost function through a regularization term, which renders unique IP solutions. Examples of model-based methods, which have been prevalent over the last two decades, include the Iterative Shrinkage and Thresholding Algorithm (ISTA)~\cite{ISTA, fast_ISTA}, Bayesian Compressive Sensing~\cite{BCS}, and Compressive Sampling Matching Pursuit~\cite{Compressive_Sensing, CoSamp}. These previous approaches can be viewed, from a Bayesian MAP perspective, as implementing a generalized Gaussian prior~\cite{fast_ISTA, ISTA, CoSamp}. Based upon the fact that the generalized Gaussian prior is subsumed by the compound Gaussian (CG) prior, which better captures statistical properties of images~\cite{Scale_Mixtures, Wavelet_Trees, CGNetTSP, DRCGNetTCI, chance2011information, waveform_opt}, the authors have previously developed an improved set of model-based methods for linear IPs utilizing a CG prior, which have shown state-of-the-art performance in tomographic imaging and CS~\cite{CGNetTSP, APSIPAlyonsrajcheney, DRCGNetTCI, HB-MAP, Asilomarlyonsrajcheney, CG_GEB}.

With advancements in artificial intelligence and deep learning (DL), a new approach to improve IP solutions is to utilize a learned prior rather than a handcrafted prior (i.e., generalized Gaussian or CG). One such approach is to create a deep neural network (DNN) by applying algorithm unrolling~\cite{learned_ISTA} to a model-based method in which a parameterization of the regularization term is learned~\cite{CGNetTSP, Asilomarlyonsrajcheney, song2023MAPUN, su2020iPianoNet, adler2018LPD, learned_proximal_operators}. Another approach, as originally proposed by Bora \textit{et al.}~\cite{ganbora}, is to employ a generative adversarial network (GAN)~\cite{goodfellow2014gans}, trained to sample from the distribution of the IP input model parameters, as the learned prior. In solving the (linear) IP, these GAN-prior approaches constrain the output from a model-based method onto the range of the GAN~\cite{ganbora, shah2018solving, dhar2018deviations, latorre2019GANADMM, song2019surfing, jalal2020robust, asim2020invertible, asim2020blind, asim2018solving, pmlr-v107-aubin20a, hand2019global, hand2018phase}. While these previous GAN-prior approaches have produced excellent empirical results and outperformed state-of-the-art model-based methods in cases where the forward measurement mapping severely undersamples the input model parameters, they suffer from a number of limitations including:
\begin{itemize}
    \item The GAN must capture a superb representation of the IP input model parameters. Typically, this is achieved by training the GAN with a substantial amount of data.
    \item Generalization to alternative datasets of IP input model parameters inherently results in significant performance loss or requires the training of another GAN.
    \item Performance, in the quality of the IP solutions, saturates with increasing sampling by the forward measurement mapping.
    \item Theoretical guarantees are often elusive, due to the non-convex GAN as a prior.
\end{itemize}
In this paper, we introduce a novel method incorporating a dual-structured prior, based on both the CG and GAN priors, to address the aforementioned four limitations.

Lastly, we remark that some previous works with GAN priors incorporate training of the GAN while solving the (linear) IP, which can boost performance and reduce saturation~\cite{jagatap2019phase, kabkab2018task, Ulyanov_2018_CVPR, mardani2019generativeMRI}. While we could incorporate such techniques into our proposed method, we focus on the case of using pretrained GAN priors and leave the case of adaptive GAN optimization (i.e., training the GAN while solving the IP) as an important topic of future work. Additionally, these adaptive GAN optimization approaches require substantial computational time and resources as a new GAN must be trained for each forward measurement mapping considered.

\subsection{Contributions}
This paper constructs an iterative estimation algorithm with a CG-GAN-prior to solve linear IPs. Specifically, we:
\begin{enumerate}
    \item Develop a novel CG-GAN-based iterative algorithm for IPs using the Alternating Direction Method of Multipliers (ADMM)~\cite{boyd2011distributed, bertsekas1976penalty, boyd2004convex}, which we denote by Alternating direction method of multipliers with Compound Gaussian and Generative Adversarial Network priors (A-CG-GAN).
    \item Derive a theoretical convergence analysis for the proposed A-CG-GAN to provide insight into the performance of the algorithm.
    \item Present ample empirical results for A-CG-GAN on CS and CT linear IPs. We demonstrate the effectiveness of A-CG-GAN even with modestly trained GANs and show that the performance of A-CG-GAN does not saturate with higher sampling from the forward measurement mapping. Furthermore, we present a clear improvement over state-of-the-art GAN-prior approaches in reconstructed image quality and generalizability.
\end{enumerate}
Our proposed CG-GAN iterative algorithm can be viewed as an expansion on~\cite{latorre2019GANADMM} by incorporating the powerful CG prior into the IP framework as in~\cite{HB-MAP, fast_HB-MAP, APSIPAlyonsrajcheney, CGNetTSP, DRCGNetTCI, CG_GEB}. By combining these two techniques we gain the advantage of using a GAN-prior while generalizing to a broader signal prior within the CG-prior context. We remark that our proposed algorithm can reduce to~\cite{latorre2019GANADMM} as a special case, which has shown excellent empirical performance when the image to be reconstructed lies in the range of the generator. As such, our proposed algorithm directly inherits such performance in the case of a GAN perfectly capturing the desired true images.

\subsection{Preliminaries}
In this section, we briefly introduce the CG prior and GANs as a prior as these are the key components we combine to furnish the novel method of this paper. First, the forward measurement mapping we consider throughout this paper is
\begin{equation}
    \bm{y} = A\bm{c} + \bm{\nu} \label{eqn:linear_msrmt}
\end{equation}
where $\bm{y}\in\mathbb{R}^m$ are the output observations (or measurements), $A\in\mathbb{R}^{m\times n}$ is a sensing matrix, $\bm{c}\in\mathbb{R}^n$ are the unknown input model parameters, and $\bm{\nu}\in\mathbb{R}^m$ is additive white noise. In many applications of interest, $m\ll n$ and $A$ is decomposed as $A = \Psi\Phi$ for $\Psi\in\mathbb{R}^{m\times n}$ a measurement matrix and $\Phi\in\mathbb{R}^{n\times n}$ a change of basis dictionary. This formulation results from representing some original input model parameters, $\bm{s}$, with respect to (w.r.t) $\Phi$ as $\bm{s} = \Phi\bm{c}$. The problem of signal reconstruction, or estimation, is the linear IP of recovering $\bm{c}$, or $\bm{s}$, given $\bm{y}$, $\Psi$, and $\Phi$.

\subsubsection{Compound Gaussian Prior}
 
A fruitful way to formulate IPs is by Bayesian estimation. In particular, the maximum a posteriori (MAP) estimate of $\bm{c}$ from (\ref{eqn:linear_msrmt}) is
\begin{align}
    \scalebox{1}{$\bm{c}^* = \arg\min_{\bm{c}}\,\, \frac{1}{2}\norm{\bm{y}-A\bm{c}}_2^2 -\log(p_{c}(\bm{c}))$}. \label{eqn:MAP estimation}
\end{align}
where $p_{c}$ is the assumed prior density of $\bm{c}$. Clearly, the prior density of $\bm{c}$, which incorporates domain-level knowledge into the IP, is crucial to a successful MAP estimate. Many previous works employ a generalized Gaussian prior~\cite{Scale_Mixtures, Wavelet_Trees} such as a Gaussian prior for Tikhonov regression~\cite{bertero1988linear} or a Laplacian prior as is predominant in the CS framework~\cite{Compressive_Sensing, ISTA, fast_ISTA, CoSamp}. 

Through the study of the statistics of image sparsity coefficients, it has been shown that coefficients of natural images exhibit self-similarity, heavy-tailed marginal distributions, and self-reinforcement among local coefficients~\cite{Wavelet_Trees}. Such properties are not encompassed by the generalized Gaussian prior. Instead, CG densities~\cite{HB-MAP}, also known as Gaussian scale mixtures~\cite{Scale_Mixtures, Wavelet_Trees}, better capture these statistical properties of natural images and images from other modalities such as radar~\cite{chance2011information, waveform_opt}. A useful formulation of the CG prior lies in modeling $\bm{c}$ as the Hadamard product
\begin{align}
    \bm{c} &= \bm{z}\odot\bm{u} \coloneqq [z_1u_1, z_2u_2, \ldots, z_nu_n]^T \label{eqn:CG}
\end{align}
such that $\bm{u}\sim \mathcal{N}(\bm{\mu}_u,\Sigma_u^{-1})$, $\bm{z}\sim p_{z}$ is a sparse or heavy-tailed positive random vector, and $\bm{u}$ and $\bm{z}$ are independent~\cite{Wavelet_Trees,HB-MAP}. We call $\bm{u}$ and $\bm{z}$ the Gaussian variable and scale variable, respectively. By suitably defining the distribution of $\bm{z}$, the CG prior subsumes many well-known distributions including the generalized Gaussian, student's $t$, $\alpha$-stable, and symmetrized Gamma distributions~\cite{Wavelet_Trees, CGNetTSP}.

We note that decomposing $\bm{c}$ as in (\ref{eqn:CG}), the joint MAP estimate of $\bm{z}$ and $\bm{u}$ from (\ref{eqn:linear_msrmt}) is
\begin{multline}
        \scalebox{1}{$\bm{z}^*, \bm{u}^* = \arg\min_{[\bm{z}, \bm{u}]}\,\, \frac{1}{2}\norm{\bm{y} - A(\bm{z}\odot\bm{u})}_2^2  - \log(p_{z}(\bm{z}))$} \\
    \scalebox{1}{$+ \frac{1}{2}(\bm{u}-\bm{\mu}_u)^T\Sigma_u (\bm{u}-\bm{\mu}_u)$} \label{eqn:MAP estimation CG}
\end{multline}
where the estimated input model parameters, $\bm{c}^*$, are then given by $\bm{c}^* = \bm{z}^*\odot\bm{u}^*.$ That is, from (\ref{eqn:MAP estimation}) we set $\bm{c} = \bm{z}\odot\bm{u}$, which, due to the independence of $\bm{z}$ and $\bm{u}$, splits the prior term $\log(p_{c}(\bm{c}))$ into the sum of a prior term for $\bm{z}$ and a prior term for $\bm{u}$.

Previously, the CG prior has been used, with a log-normal distributed scale variable, to estimate images through an iterative MAP estimation~\cite{HB-MAP, fast_HB-MAP, CGNetTSP, APSIPAlyonsrajcheney} and an algorithm unrolled DNN~\cite{CGNetTSP, APSIPAlyonsrajcheney}. Additionally, a learned scale variable distribution in the CG prior was used to estimate images through an alternative algorithm unrolled DNN~\cite{DRCGNetTCI, Asilomarlyonsrajcheney}. Finally, the CG prior, using a real-valued random scale variable, has been successfully used for image denoising~\cite{portilla2003image} and hyperspectral image CS~\cite{huang2021deep}.

\subsubsection{GANs as a Prior}
To illustrate the usage of GANs as a prior in IPs, we overview the seminal work of Bora \textit{et al.}~\cite{ganbora}. Let $G:\mathbb{R}^d\to\mathbb{R}^n$ be the generative neural network (NN) from a GAN, which we assume has been adversarially trained against a discriminator NN $D:\mathbb{R}^n\to\mathbb{R},$ to sample from some distribution (e.g., $G$ produces ``natural" images). Often $d\ll n$, implying that the generative DNN maps from a low-dimensional latent space to a high-dimensional space of interest. Given a low-dimensional latent distribution, $p_x$, over $\mathbb{R}^d$, most commonly $p_x\sim \mathcal{N}(0,\sigma^2 I)$, we can generate a distribution from $G$ as $p_G \sim G(\bm{x})$ for $\bm{x}\sim p_x$.

A generative NN is incorporated as a prior by assuming that $\bm{c}$ from (\ref{eqn:linear_msrmt}), satisfies $\bm{c}\sim p_G.$ That is, there exists a latent vector $\bm{x}\sim p_x$ such that $\bm{c} = G(\bm{x}).$ Therefore, rather than optimize for the solution to the IP directly as in (\ref{eqn:MAP estimation}), we optimize over the latent space to find the best $\bm{x}^*$ such that $G(\bm{x}^*)\approx \bm{c}$. On this front, consider the MAP estimate of $\bm{x}$ from (\ref{eqn:linear_msrmt}), when $\bm{c} = G(\bm{x})$, given by
\begin{align}
    \scalebox{1}{$\bm{x}^* = \arg\min_{\bm{x}}\,\, \frac{1}{2}\norm{\bm{y}-AG(\bm{x})}_2^2 + \frac{1}{2\sigma^2}\norm{\bm{x}}_2^2.$} \label{eqn:MAP estimation GAN}
\end{align}
Intuitively, if $\bm{c}$ is a ``natural" image and $G$ samples from the distribution of ``natural" images then we can expect that $\bm{c} = G(\bm{x})$ for some $\bm{x}\sim p_x.$

\section{ADMM with CG-GAN Prior (A-CG-GAN)} \label{sec:iterative_alg}

To solve the linear IP to (\ref{eqn:linear_msrmt}) for the unknown vector $\bm{c}$, we incorporate both the CG prior, as in (\ref{eqn:MAP estimation CG}), and GAN prior, as in (\ref{eqn:MAP estimation GAN}), by setting the scale variable to lie within the range of a generative NN. Mathematically, we consider the following constrained optimization problem
\begin{align}
    (\bm{x}^*, \bm{z}^*, \bm{u}^*,  \bm{c}^*) &= \underset{(\bm{x},\bm{z},\bm{u},\bm{c})\in\mathcal{D}_x\times\mathcal{D}_z\times\mathcal{D}_u\times\mathcal{D}_c}{\arg\min}\,\, F(\bm{x}, \bm{z},\bm{u},\bm{c}) \nonumber  \\
  &\hspace{.4cm} \text{s.t.} \s \bm{z} =  G(\bm{x}), \hspace{.5cm} \bm{c} = \bm{z}\odot\bm{u} \label{eqn:iterative_opt}
    \end{align}
    where
\begin{multline}
       \scalebox{1}{$F(\bm{x}, \bm{z},\bm{u},\bm{c}) \coloneqq \frac{1}{2}\norm{\bm{y}-A\bm{c}}_2^2 + \mu \norm{\bm{z}}_1  + R_x(\bm{x})$} \\
     \scalebox{1}{$+ \frac{\lambda}{2}(\bm{u}-\bm{\mu}_u)^T\Sigma_u(\bm{u}-\bm{\mu}_u),$} \label{eqn:cost function} 
\end{multline}
$G:\mathbb{R}^d\to\mathbb{R}^n$ is a generative NN, $R_x:\mathbb{R}^d\to\mathbb{R}$ is a convex regularization function, $\mu$ and $\lambda$ are positive scaling parameters, and $\mathcal{D}_x, \mathcal{D}_z, \mathcal{D}_u,$ and $\mathcal{D}_c$ are the convex domains of $\bm{x}, \bm{z}, \bm{u}$, and $\bm{c}$, respectively. We remark that the cost function, $F$, consists of a data fidelity term, a regularization term enforcing sparsity of $\bm{z}$, an implicit regularization term on $\bm{x}$, and a regularization term enforcing Gaussianity of $\bm{u}$.

We utilize ADMM~\cite{boyd2011distributed, bertsekas1976penalty, boyd2004convex} to solve the constrained minimization problem of (\ref{eqn:iterative_opt}). For this, define the feasibility gap
\begin{align}
    \bm{\xi} \equiv \bm{\xi}(\bm{x},\bm{z},\bm{u},\bm{c}) \coloneqq \begin{bmatrix}
        (\bm{z}-G(\bm{x}))^T & (\bm{c}-\bm{z}\odot\bm{u})^T
    \end{bmatrix}^T \label{eqn:feasibility gap}
\end{align}
and let $\bm{\xi}_k \equiv \bm{\xi}(\bm{x}_k,\bm{z}_k,\bm{u}_k,\bm{c}_k)$. Additionally, define the augmented Lagrangian
\begin{align}
    \scalebox{1}{$\mathcal{L}_{\rho}(\bm{x},\bm{z},\bm{u},\bm{c},\pmb{\phi}) \coloneqq F(\bm{x},\bm{z},\bm{u},\bm{c}) + \left\langle \pmb{\phi}, \bm{\xi} \right\rangle + \frac{\rho}{2}\norm{\bm{\xi}}_2^2 $}\label{eqn:augmented Lagrangian}
\end{align}
for a positive scalar $\rho$ and a dual variable $\pmb{\phi}\in\mathbb{R}^{2n}$. Therefore, the optimization problem of (\ref{eqn:iterative_opt}) is equivalent to
\begin{align}
    \underset{(\bm{x},\bm{z},\bm{u},\bm{c})\in\mathcal{D}_x\times\mathcal{D}_z\times\mathcal{D}_u\times\mathcal{D}_c}{\arg\min}\,\,\, \underset{\pmb{\phi}}{\arg\max}\,\, \mathcal{L}_{\rho}(\bm{x},\bm{z},\bm{u},\bm{c},\pmb{\phi}), \label{eqn:iterative_opt_equiv}
\end{align}
and our A-CG-GAN algorithm iteratively solves (\ref{eqn:iterative_opt_equiv}) through an ADMM block coordinate descent~\cite{wright2015coordinate} as given in Algorithm~\ref{alg:CG-GAN}.

To detail Algorithm~\ref{alg:CG-GAN}, let $\pmb{\phi} = \begin{bmatrix}
        \bm{\varphi}_1^T & \bm{\varphi}_2^T
    \end{bmatrix}^T$ and define
\begin{align}
 \mathcal{T}_c &\equiv \mathcal{T}_c(\bm{z},\bm{u},\pmb{\phi}) \coloneqq \underset{\bm{c}}{\arg\min}\,\,  \mathcal{L}_{\rho}(\bm{x},\bm{z},\bm{u},\bm{c},\pmb{\phi}) \nonumber \\
 &= (A^TA+\rho I)^{-1}(A^T\bm{y}+\rho \bm{z}\odot\bm{u} - \bm{\varphi}_2) \label{eqn:Tik c} \\
 \mathcal{T}_u &\equiv \mathcal{T}_u(\bm{z},\bm{c},\pmb{\phi}) \coloneqq \underset{\bm{u}}{\arg\min}\,\,  \mathcal{L}_{\rho}(\bm{x},\bm{z},\bm{u},\bm{c},\pmb{\phi}) \nonumber  \\
 &\scalebox{0.99}{${\displaystyle = (\rho\text{Diag}(\bm{z})^2 + \lambda \Sigma_u)^{-1}(\lambda \Sigma_u \bm{\mu}_u + \bm{z}\odot(\rho\bm{c} + \bm{\varphi}_2)).}$} \label{eqn:Tik u}
\end{align}
As the feasibility terms $\left\langle \pmb{\phi}, \bm{\xi} \right\rangle + \frac{\rho}{2}\norm{\bm{\xi}}_2^2$ in (\ref{eqn:augmented Lagrangian}) are convex w.r.t $\bm{z}$, we minimize $\mathcal{L}_{\rho}$ in (\ref{eqn:augmented Lagrangian}) w.r.t~$\bm{z}$ using a standard fast ISTA (FISTA)~\cite{fast_ISTA} technique. We will write
\begin{align*}
    \mathcal{T}_z^{(J)} \equiv \mathcal{T}_z^{(J)}(\bm{x},\bm{z},\bm{u},\bm{c},\pmb{\phi}) = J \text{ FISTA steps on } \mathcal{L}_{\rho} \text{ w.r.t } \bm{z}.
\end{align*}

\begin{algorithm}[!t]
\caption{ADMM with CG-GAN Prior (A-CG-GAN)}\label{alg:CG-GAN}
\begin{algorithmic}[1]
\REQUIRE $\bm{y}, A, G, \mu, \lambda, \bm{\mu}_u, R_x, K, J, J_x$ and tolerance parameter $\tau$
\STATE $\bm{x}_0 \sim \mathcal{N}(\bm{0}, I)$, $\bm{z}_0 =  G(\bm{x}_0)$, $\bm{u}_0 = \mathcal{T}(\bm{z}_0)$, $\bm{c}_0 = \bm{z}_0\odot\bm{u}_0$, and $\pmb{\phi}_0 = \bm{0}$
\FOR{$k\in \{0,1,\ldots, K\}$}
\STATE  $\bm{x}_{k+1} = \mathcal{T}_x^{(J_x)}(\bm{x}_k, \bm{z}_k, \pmb{\phi}_k)$ \label{line:x update}
\STATE  $\bm{z}_{k+1} = \mathcal{T}_z^{(J)}(\bm{x}_{k+1}, \bm{z}_k, \bm{u}_k, \bm{c}_k, \pmb{\phi}_k)$ \label{line:z update}
\STATE  $\bm{u}_{k+1} = \mathcal{T}_u(\bm{z}_{k+1}, \bm{c}_k, \pmb{\phi}_k)$
\STATE $\bm{c}_{k+1} = \mathcal{T}_c(\bm{z}_{k+1},\bm{u}_{k+1}, \pmb{\phi}_k) $ 
\STATE $\pmb{\phi}_{k+1} = \pmb{\phi}_k + \sigma_{k+1} \bm{\xi}_{k+1}$ \label{line:dual update}
\IF{$ \frac{1}{d}\delta_{x,k}^2+ \frac{1}{n}\left(\delta_{z,k}^2 + \delta_{u,k}^2 + \delta_{c,k}^2 + \norm{\bm{\xi}_{k+1}}_2^2\right) < \tau$}
\RETURN $(\bm{x}_{k+1}, \bm{z}_{k+1}, \bm{u}_{k+1}, \bm{c}_{k+1})$
\ENDIF
\ENDFOR
\ENSURE $(\bm{x}_K, \bm{z}_K, \bm{u}_K, \bm{c}_K)$
\end{algorithmic}
\end{algorithm}

Next, for the update of $\bm{x}$, which requires optimizing over the non-convex generative NN $G$, we, similar to~\cite{ganbora, dhar2018deviations, shah2018solving}, use adaptive moment estimation (Adam)~\cite{ADAM}, a gradient descent (GD)-based optimizer, with a step size of $10^{-2}.$ The gradient, $\nabla_{\bm{x}} \mathcal{L}_{\rho}$, for each GD step is calculated using backpropagation and automatic differentiation~\cite{auto_diff} in TensorFlow~\cite{abadi2016tensorflow}. We will write
\begin{align*}
    \mathcal{T}_x^{(J_x)} \equiv \mathcal{T}_z^{(J_x)}(\bm{x},\bm{z},\pmb{\phi})  = J_x \text{ GD steps on } \mathcal{L}_{\rho} \text{ w.r.t } \bm{x}.
\end{align*}
As this update requires $R_x$ to be differentiable, we instead use $J_x$ ISTA or proximal gradient descent (PGD) steps to minimize $\mathcal{L}_{\rho}$ w.r.t $\bm{x}$ when $R_x$ is non-smooth and also denote these $J_x$ steps as $\mathcal{T}_x^{(J_x)}$ for simplicity.

The dual variable, $\pmb{\phi}$, is updated via a single gradient ascent step on $\mathcal{L}_{\rho}$ as is typical in ADMM~\cite{boyd2011distributed, bertsekas1976penalty, boyd2004convex}. That is,
\begin{align*}
    \pmb{\phi} \leftarrow \pmb{\phi} + \sigma \nabla_{\pmb{\phi}} \mathcal{L}_{\rho} = \pmb{\phi} + \sigma \bm{\xi}
\end{align*}
where $\sigma$ is a real-value step size parameter.

As specified in Algorithm~\ref{alg:CG-GAN}, the initial estimates are given by $\bm{x}_0 \sim \mathcal{N}(\bm{0}, I)$, $\bm{z}_0 =  G(\bm{x}_0)$, $\bm{u}_0 = \mathcal{T}(\bm{z}_0)$, and $\bm{c}_0 = \bm{z}_0\odot\bm{u}_0$ where, for $A_{\bm{z}} = A\text{Diag}(\bm{z})$,
\begin{align*}
    \resizebox{\columnwidth}{!}{$\mathcal{T}(\bm{z}) \coloneqq \underset{\bm{u}}{\arg\min}\,\, \widehat{F}(\bm{z},\bm{u}) = (A_{\bm{z}}^TA_{\bm{z}} + \lambda\Sigma_u)^{-1}(\lambda\Sigma_u\bm{\mu}_u + A_{\bm{z}}^T\bm{y})$}
\end{align*}
for $\widehat{F}(\bm{z},\bm{u}) = \frac{1}{2}\norm{\bm{y}-A(\bm{z}\odot\bm{u})}_2^2 + \frac{\lambda}{2}(\bm{u}-\bm{\mu}_u)^T\Sigma_u(\bm{u}-\bm{\mu}_u)$. This particular choice of $\bm{u}_0$ is inspired by previous work in regularizing IPs with a CG prior~\cite{CGNetTSP, DRCGNetTCI}.

Lastly, define $\delta_{x,k} = \norm{\bm{x}_{k+1}-\bm{x}_k}_2$ and similarly define $\delta_{z,k}, \delta_{u,k},$ and $\delta_{c,k}$. Convergence of Algorithm~\ref{alg:CG-GAN} is determined by $\frac{1}{d}\delta_{x,k}^2+ \frac{1}{n}\left(\delta_{z,k}^2 + \delta_{u,k}^2 + \delta_{c,k}^2 + \norm{\bm{\xi}_{k+1}}_2^2\right) < \tau$ for a tolerance parameter $\tau$. This stopping condition, informed by previous ADMM work~\cite{boyd2011distributed, latorre2019GANADMM}, ensures that the mean squared error in both the feasibility gap and the change in each of the four primal variables is sufficiently small.

We make a few remarks about Algorithm~\ref{alg:CG-GAN}. First, (\ref{eqn:Tik c}) and (\ref{eqn:Tik u}) can be quickly calculated using SVD. Specifically, if $A = USV^T$ is the SVD of $A$, then (\ref{eqn:Tik c}) can be rewritten as
\begin{align*}
    \mathcal{T}_c &= V(S^TS+\rho I)^{-1}V^T(A^T\bm{y}+\rho \bm{z}\odot\bm{u} - \bm{\varphi}_2),
\end{align*}
which is easily calculated since $S^TS+\rho I$ is a diagonal matrix. Similarly, for (\ref{eqn:Tik u}), since $\Sigma_u$ is symmetric and positive definite then the SVD of $\Sigma_u$ has the form $\Sigma_u = VSV^T$, which will reduce the matrix inverse in (\ref{eqn:Tik u}) to inverting a diagonal matrix. Therefore, by performing the SVD of $A$ and $\Sigma_u$ upfront, only diagonal matrices need to be inverted during the iterations of Algorithm~\ref{alg:CG-GAN}, which provides a considerable time reduction.

Finally, as A-CG-GAN expands on the methodology introduced by~\cite{latorre2019GANADMM}, we briefly explain the core difference. In~\cite{latorre2019GANADMM}, only a single constraint from the GAN prior, specifically $\bm{c} = G(\bm{x})$, is imposed. Accordingly, each iteration of the ADMM algorithm in~\cite{latorre2019GANADMM} consists of two primal variable updates for $\bm{c}$ and $\bm{x}$, similar to lines \ref{line:x update} and \ref{line:z update} in Algorithm~\ref{alg:CG-GAN}, and one dual variable update $\pmb{\phi}\leftarrow \pmb{\phi} + \sigma (\bm{c}_{k+1}-G(\bm{x}_{k+1}))$, similarly to line \ref{line:dual update} in Algorithm~\ref{alg:CG-GAN}. In our proposed A-CG-GAN, we have a similar GAN prior constraint now involving the generator, $G$, and the scale variable, $\bm{z}$, as we constrain $\bm{z}$ to be in the range of $G$. Uniquely, we introduce a second constraint to additionally enforce the CG prior by constraining $\bm{c}$ to be the product of a scale variable, $\bm{z}$, and Gaussian variable, $\bm{u}$.

\section{Convergence Results}
\subsection{Preliminaries}
For our theoretical analysis in this section and the Appendix, we will use the following notation and nomenclature:
\begin{enumerate}
    \item $\bm{\zeta}_1 \coloneqq \bm{x}$, $\bm{\zeta}_2 \coloneqq \bm{z}$, $\bm{\zeta}_3\coloneqq \bm{u}$, and $\bm{\zeta}_4 \coloneqq \bm{c}$
    \item $\bm{\zeta}_1^{(k)} \coloneqq \bm{x}_k$, $\bm{\zeta}_2^{(k)} \coloneqq \bm{z}_k$, $\bm{\zeta}_3^{(k)} \coloneqq \bm{u}_k$, and $\bm{\zeta}_4^{(k)} \coloneqq \bm{c}_k$
    \item $\bm{\zeta}_1^*\coloneqq \bm{x}^*$, $ \bm{\zeta}_2^* \coloneqq \bm{z}^*$, $\bm{\zeta}_3^*\coloneqq \bm{u}^*$, and $\bm{\zeta}_4^*\coloneqq\bm{c}^*$
    \item $\bm{v} \coloneqq [\bm{\zeta}_i]_{i = 1}^4$, $\bm{v}_k \coloneqq [\bm{\zeta}_i^{(k)}]_{i = 1}^4$, and $\bm{v}^* \coloneqq [\bm{\zeta}_i^{*}]_{i = 1}^4$
    \item  $\bm{v}_{i_1:i_2} \coloneqq [\bm{\zeta}_i]_{i = i_1}^{i_2}$ and  $\bm{v}_{{i_1:i_2}}^{(k)} \coloneqq [\bm{\zeta}_i^{(k)}]_{i = i_1}^{i_2}$
    \item $\Delta_{i,k} \coloneqq \max\left\{\norm{\bm{\zeta}_i^* - \bm{\zeta}_i^{(k)}}_2,\,\, \norm{\bm{\zeta}_i^* - \bm{\zeta}_i^{(k+1)}}_2\right\}$
    \item $\Delta_{(2,3), k} \coloneqq \norm{\bm{z}^*\odot\bm{u}^* - \bm{z}_k\odot\bm{u}_k}_2$
    \item $\delta_{i,k} \coloneqq \norm{\bm{\zeta}_i^{(k+1)} - \bm{\zeta}_i^{(k)}}_2$
    \item $\pmb{\phi}_k \coloneqq \begin{bmatrix}
        \bm{\varphi}_{1, k}^T& \bm{\varphi}_{2, k}^T
    \end{bmatrix}^T$ and $\pmb{\phi}^* \coloneqq \begin{bmatrix}
        (\bm{\varphi}_{1}^*)^T& (\bm{\varphi}_{2}^*)^T
    \end{bmatrix}^T$ 
    \item $\mathcal{L}_k \coloneqq \mathcal{L}_{\rho}(\bm{v}^{(k)}_{1:4}, \pmb{\phi}_k)$ and $\mathcal{L}_* \coloneqq \mathcal{L}_{\rho}(\bm{v}^*, \pmb{\phi}^*)$ 
    \item $\overline{\mathcal{L}}_k \coloneqq \overline{\mathcal{L}}_{\rho}(\bm{v}^{(k)}_{1:4}, \pmb{\phi}_k)$ and $\overline{\mathcal{L}}_* \coloneqq \overline{\mathcal{L}}_{\rho}(\bm{v}^*,\pmb{\phi}^*)$ for $\overline{\mathcal{L}}_{\rho}$ defined in (\ref{eqn:augmented Lagrangian generalized adjusted}) 
    \item $\bm{\xi}_{1,k} \equiv \bm{\xi}_{1,k}(\bm{x}_k,\bm{z}_k) \coloneqq \bm{z}_k - G(\bm{x}_k)$
    \item $\bm{\xi}_{2,k} \equiv \bm{\xi}_{2,k}(\bm{z}_k,\bm{u}_k,\bm{c}_k) \coloneqq \bm{c}_k - \bm{z}_k\odot\bm{u}_k$
    \item $\bm{\xi}_k \equiv \bm{\xi}_k(\bm{x}_k,\bm{z}_k,\bm{u}_k,\bm{c}_k) \coloneqq \begin{bmatrix}
        \bm{\xi}_{1,k}^T & \bm{\xi}_{2,k}^T
    \end{bmatrix}^T.$
    \item $\mathcal{B}^n_{t} \coloneqq \{\bm{t}\in\mathbb{R}^n : \norm{\bm{t}}_2\leq t\}$
\end{enumerate}
Additionally, we will generalize the cost function in (\ref{eqn:cost function}) to
\begin{align}
    F(\bm{v}) \coloneqq f_{\bm{y}}(\bm{c}) + R_1(\bm{x}) + R_2(\bm{z}) + R_3(\bm{u}) + R_4(\bm{c}) \label{eqn:iterative cost fnc generalized}
\end{align}
where $f_{\bm{y}}:\mathbb{R}^n\to\mathbb{R}$ is a data fidelity term and $R_1, R_2, R_3,$ and $R_4$ are regularization functions on $\bm{x}, \bm{z}, \bm{u},$ and $\bm{c}$, respectively. We make the following assumptions about the terms of (\ref{eqn:iterative cost fnc generalized}).
\begin{assumption} \label{assume:cost function}
    Each function $f_{\bm{y}}$ and $R_i$ is continuous, convex, and has Lipschitz continuous gradient with constant $L_f$ and $L_i$, respectively.
\end{assumption}

We define $\mathcal{L}_{\rho}$ as in (\ref{eqn:augmented Lagrangian}) with $F$ now given in (\ref{eqn:iterative cost fnc generalized}) and define an adjusted augmented Lagrangian as
\begin{align}
    \overline{\mathcal{L}}_{\rho}(\bm{v}, \pmb{\phi}) \coloneqq f_{\bm{y}}(\bm{c}) + \langle \pmb{\phi}, \bm{\xi}\rangle + \frac{\rho}{2}\norm{\bm{\xi}}_2^2. \label{eqn:augmented Lagrangian generalized adjusted}
\end{align}
That is, $\overline{\mathcal{L}}_{\rho}$ is the augmented Lagrangian without the regularization terms. Despite the generalization of the cost function, we assume that, as holds in Section~\ref{sec:iterative_alg}, the minimizer of $\mathcal{L}_{\rho}$ w.r.t each $\bm{\zeta}_i$ individually is well approximated by GD/ISTA/FISTA steps (or for $\bm{u}$ and $\bm{c}$ is analytically known).

Note that we set $\mathcal{D}_x = \mathcal{B}_{x_\infty}^d$, $\mathcal{D}_z = \mathcal{B}_{z_\infty}^n$, $\mathcal{D}_u = \mathcal{B}_{u_\infty}^n$, and $\mathcal{D}_c = \mathcal{B}_{c_\infty}^n$. Furthermore, to prove convergence, we will take an adaptive dual-variable step size of
\begin{align}
     \scalebox{0.99}{${\displaystyle\sigma_{k} = \min\left\{\sigma_{0}, \frac{\sigma_{0}}{\max\left\{\norm{\bm{\xi}_{1,k}}_2, \norm{\bm{\xi}_{2,k}}_2 \right\} k\ln^2(k+1)}\right\},}$} \label{eqn:sigma_k for theory}
\end{align}
for some initial step size $\sigma_0$, which is utilized in~\cite{latorre2019GANADMM} and motivated by~\cite{bertsekas1976penalty} as a guarantee on maintaining a bounded dual-variable. While such a step size is chosen to provide theoretical guarantees, we find, in practice, that a constant step size $\sigma_{k+1} = \rho \leq 1$ is sufficient for bounded dual-variables.

Lastly, as our results build off of the convergence properties discussed in~\cite{latorre2019GANADMM}, which again considers an ADMM algorithm with a single GAN prior constraint, we import necessary assumptions on $G$, the generative NN, from~\cite{latorre2019GANADMM}.
\begin{assumption} \label{assume:G}
    For any $\widehat{\bm{x}}, \widetilde{\bm{x}}\in\mathbb{R}^d$ the following bounds hold
    \begin{align*}
        \nu_G \norm{\widehat{\bm{x}} - \widetilde{\bm{x}}}_2\leq \norm{G(\widehat{\bm{x}}) - G(\widetilde{\bm{x}})}_2 &\leq \tau_G\norm{\widehat{\bm{x}} - \widetilde{\bm{x}}}_2 \\
        \norm{G(\widehat{\bm{x}}) - G(\widetilde{\bm{x}}) - (\nabla_{\bm{x}} G(\widetilde{\bm{x}}))^T(\widehat{\bm{x}}-\widetilde{\bm{x}})}_2 &\leq \frac{L_G}{2}\norm{\widehat{\bm{x}}-\widetilde{\bm{x}}}_2^2
    \end{align*}
    for positive constants $\nu_G, \tau_G,$ and $L_G$.
\end{assumption}
We remark that the first inequality in Assumption~\ref{assume:G} implies that $G$ is a near-isometry while the second inequality is a strong smoothness requirement on $G$. See~\cite{latorre2019GANADMM} for more details about generative NNs that satisfy Assumption~\ref{assume:G}.

\subsection{Core Theoretical Guarantees} \label{sec:core theory}
In this section, we provide our main technical results for the convergence of our generalized A-CG-GAN algorithm, which are inspired by the previous work in~\cite{latorre2019GANADMM}. For this, we let $\bm{v}^*$ be some solution of (\ref{eqn:iterative_opt}), with $F$ given in (\ref{eqn:iterative cost fnc generalized}), and $\pmb{\phi}^*$ be the corresponding optimal dual variable from (\ref{eqn:iterative_opt_equiv}). 

First, we present four propositions bounding the augmented Lagrangian change over an update of each primal variable.
\begin{proposition}\label{prop:w update bound}
    Let $\alpha_4 = \frac{2\alpha_c}{1+2\alpha_c}$ for any $\alpha_c\leq (L_f+\rho)^{-1}$. Then for any $\theta\in [0,1]$ and $k\in\mathbb{N}$ the following bound holds
    \begin{align*}
            &\overline{\mathcal{L}}_{\rho}(\bm{v}_{1:4}^{(k+1)}, \pmb{\phi}_k) + R_4(\bm{c}_{k+1}) \\
            &\leq \overline{\mathcal{L}}_{\rho}(\bm{v}_{1:3}^{(k+1)}, \bm{c}_k, \pmb{\phi}_k) + \theta\langle \bm{c}^*-\bm{c}_k, \nabla_{\bm{c}} \overline{\mathcal{L}}_{k}\rangle  + \frac{\theta^2}{\alpha_4} \Delta_{4,k}^2  \\
    &\s + \frac{(\rho u_{\infty})^2}{2} \delta_{2,k}^2  + \frac{(\rho z_{\infty})^2}{2}\delta_{3,k}^2 + \theta R_4(\bm{c}^*) + (1-\theta)R_4(\bm{c}_k). 
    \end{align*}
\end{proposition}

\begin{proposition}\label{prop:u update bound}
    Let $\alpha_3 = \frac{2\alpha_u}{1+\alpha_u}$ for any $\alpha_u\leq (\rho z_{\infty}^2)^{-1}.$ Then for any $\theta\in [0,1]$ and $k\in\mathbb{N}$ there exists a $\gamma \geq 0$ such that
    \begin{align*}
        &\overline{\mathcal{L}}_{\rho}(\bm{v}_{1:3}^{(k+1)}, \bm{c}_k, \pmb{\phi}_k) + R_3(\bm{u}_{k+1})\\
        &\leq \overline{\mathcal{L}}_{\rho}(\bm{v}_{1:2}^{(k+1)}, \bm{v}_{3:4}^{(k)}, \pmb{\phi}_k) + \theta\langle \bm{u}^* - \bm{u}_k, \nabla_{\bm{u}} \overline{\mathcal{L}}_k\rangle  + \frac{\theta^2}{\alpha_3} \Delta_{3,k}^2  \\
        &  + \frac{\gamma}{2} \delta_{2,k}^2 + \theta R_3(\bm{u}^*) + (1-\theta)R_3(\bm{u}_k).
    \end{align*}
\end{proposition}

\begin{proposition}\label{prop:z update bound}
    Let $\alpha_2 = \frac{2\alpha_z}{1+\alpha_z}$ for any $\alpha_z\leq (\rho(1+u_{\infty}^2))^{-1}.$ Then for any $\theta\in [0,1]$ and $k\in\mathbb{N}$ the following bound holds
    \begin{align*}
        & \overline{\mathcal{L}}_{\rho}(\bm{v}_{1:2}^{(k+1)}, \bm{v}_{3:4}^{(k)}, \pmb{\phi}_k) + R_2(\bm{z}_{k+1}) \\
        & \leq \overline{\mathcal{L}}_{\rho}(\bm{x}_{k+1},\bm{v}_{2:4}^{(k)}, \pmb{\phi}_k) + \theta\langle \bm{z}^* - \bm{z}_k, \nabla_{\bm{z}} \overline{\mathcal{L}}_k\rangle  + \frac{\theta^2}{\alpha_2}\Delta_{2,k}^2  \\
        &\s + \frac{(\rho\tau_G)^2}{2}\delta_{1,k}^2  + \theta R_2(\bm{z}^*) + (1-\theta)R_2(\bm{z}_k).
    \end{align*}
\end{proposition}

\begin{proposition}\label{prop:x update bound}
    For $k\in\mathbb{N}$, let $\alpha_1 \leq  2(L_G(4\sigma_0+\rho \norm{\bm{\xi}_{1,k}}_2)+\rho\tau_G^2)^{-1}$. Then for any $\theta\in [0,1]$ the following bound holds  
    \begin{multline*}
        \overline{\mathcal{L}}_{\rho}(\bm{x}_{k+1},\bm{v}_{2:4}^{(k)}, \pmb{\phi}_k) + R_1(\bm{x}_{k+1}) \\
        \leq \overline{\mathcal{L}}_k + \theta\langle \bm{x}^*-\bm{x}_k, \nabla_{\bm{x}} \overline{\mathcal{L}}_k\rangle + \frac{\theta^2}{\alpha_1}\Delta_{1,k}^2 \\
        + \theta R_1(\bm{x}^*) + (1-\theta) R_1(\bm{x}_k).
    \end{multline*}
\end{proposition}
Proofs of propositions~\ref{prop:w update bound},~\ref{prop:u update bound},~\ref{prop:z update bound}, and~\ref{prop:x update bound} are provided respectively in Appendices~\ref{apndx:proof of prop:w update bound},~\ref{apndx:proof of prop:u update bound},~\ref{apndx:proof of prop:z update bound}, and~\ref{apndx:proof of prop:x update bound}. 

Next, we combine the previous four propositions to bound the augmented Lagrangian change over one iteration of A-CG-GAN.
\begin{proposition}\label{prop:all variable update + full inner product bound}
    For any $\theta\in [0,1]$ and every $k\in\mathbb{N}$ there exists positive constants $\{(\alpha_j, \beta_j)\}_{j = 1}^4$ such that
    \begin{align*}
        \mathcal{L}_{k+1} \leq \mathcal{L}_k + \sum_{j = 1}^4 \beta_{j} \Delta_{j,k}^2 + \theta^2\sum_{j = 1}^4 \frac{\Delta_{j,k}^2}{\alpha_j} + \theta (\mathcal{L}_* - \mathcal{L}_k).
    \end{align*}
\end{proposition}
Proposition~\ref{prop:all variable update + full inner product bound}, which provides a bound on the change in the augmented Lagrangian, is proved in Appendix~\ref{apndx:proof of prop:all variable update + full inner product bound}.

Finally, we present our main convergence theorem.

\begin{theorem}\label{thm:convergence}
     Let the assumptions of Lemma~\ref{lemma:theta choice prelim 3} hold and set $\{\kappa_j\}_{j=1}^4$ and $\mu$ as given in Lemma~\ref{lemma:theta choice prelim 3}.
    Choose $\{(\alpha_j, \beta_j)\}_{j=1}^4$, from Proposition \ref{prop:all variable update + full inner product bound}, to satisfy $ \widehat{\beta} \leq \frac{\widetilde{\beta}}{4} \leq \frac{3}{4}$ where $\widehat{\beta} \coloneqq \sqrt{\max_{1\leq i\leq 4}\,\,\alpha_i\beta_i}$ and $\widetilde{\beta} \coloneqq \min_{1\leq i\leq 4}\,\, \frac{\alpha_i\kappa_i}{4}$. Define $\widehat{\eta} \coloneqq 4\mu(\rho \widetilde{\beta})^{-1}$ and let $\sum_{j = 1}^4 \frac{\Delta_{j,k}^2}{\alpha_j} \geq \widehat{\eta}.$ Then for $\delta \coloneqq \frac{\widetilde{\beta}}{8} \in (0,1)$
    \begin{align*}
        \sum_{j = 1}^4 \frac{\Delta_{j,k}^2}{\alpha_j} \leq (1-\delta)^k \frac{\mathcal{L}_0-\mathcal{L}_*}{16\widehat{\beta}} + \frac{\widehat{\eta}}{16}.
    \end{align*}
\end{theorem}
A proof of Theorem~\ref{thm:convergence} is provided in Appendix~\ref{apndx:proof of thm:convergence}. We remark that Theorem~\ref{thm:convergence} states that outside of a neighborhood about some solution $\bm{v}^*$, our A-CG-GAN algorithm will converge, at worst, linearly with rate $(1-\delta)^k$ to this neighborhood. As the size of the neighborhood is proportional to $\widehat{\eta}$, which in turn is inversely proportional to $\rho$, we can decrease the size of the neighborhood by increasing the augmented Lagrangian penalty parameter $\rho$.

\section{Empirical Results}

\begin{figure*}[!b]
\centering
\begin{subfigure}{0.24\textwidth}
    \centering
    \includegraphics[width=\textwidth]{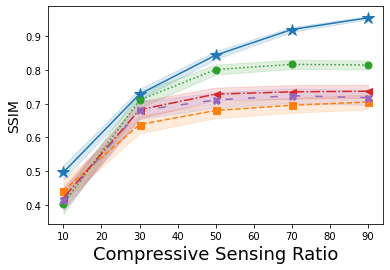}
    \caption{CIFAR10 estimates from CS with 60dB SNR.}
    \label{fig:compare_CIFAR10_ratio_60}
\end{subfigure}
\begin{subfigure}{0.24\textwidth}
    \centering
    \includegraphics[width=\textwidth]{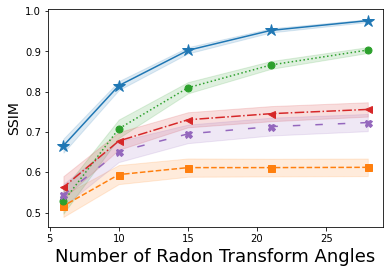}
    \caption{CIFAR10 estimates from CT with 60dB SNR.}
    \label{fig:compare_CIFAR10_angles_60}
\end{subfigure}
\begin{subfigure}{0.24\textwidth}
    \centering
    \includegraphics[width=\textwidth]{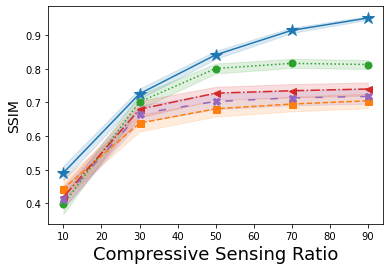}
    \caption{CIFAR10 estimates from CS with 40dB SNR.}
    \label{fig:compare_CIFAR10_ratio_40}
\end{subfigure}
\begin{subfigure}{0.24\textwidth}
    \centering
    \includegraphics[width=\textwidth]{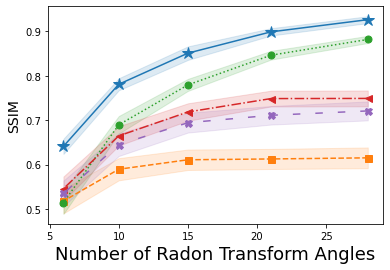}
    \caption{CIFAR10 estimates from CT with 40dB SNR.}
    \label{fig:compare_CIFAR10_angles_40}
\end{subfigure}
\begin{subfigure}{0.24\textwidth}
    \centering
    \includegraphics[width=\textwidth]{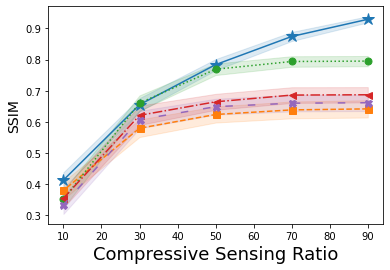}
    \caption{CalTech101 estimates from CS with 60dB SNR.}
    \label{fig:compare_CalTech101_ratio_60}
\end{subfigure}
\begin{subfigure}{0.24\textwidth}
    \centering
    \includegraphics[width=\textwidth]{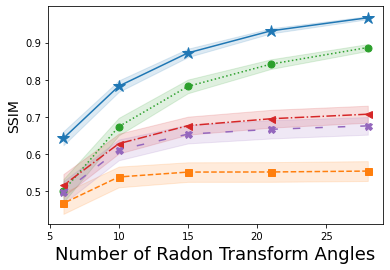}
    \caption{CalTech101 estimates from CT with 60dB SNR.}
    \label{fig:compare_CalTech101_angles_60}
\end{subfigure}
\begin{subfigure}{0.24\textwidth}
    \centering
    \includegraphics[width=\textwidth]{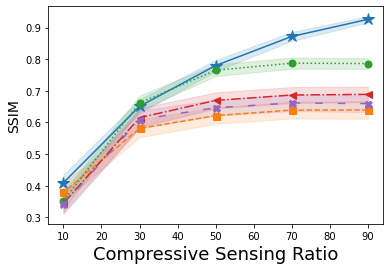}
    \caption{CalTech101 estimates from CS with 40dB SNR.}
    \label{fig:compare_CalTech101_ratio_40}
\end{subfigure}
\begin{subfigure}{0.24\textwidth}
    \centering
    \includegraphics[width=\textwidth]{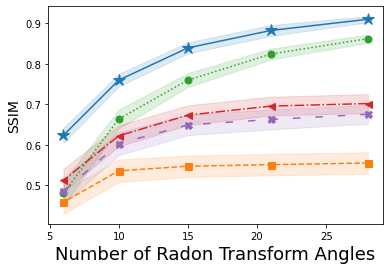}
    \caption{CalTech101 estimates from CT with 40dB SNR.}
    \label{fig:compare_CalTech101_angles_40}
\end{subfigure}
\begin{subfigure}{0.48\textwidth}
    \centering
    \includegraphics[width = \textwidth]{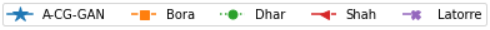}
\end{subfigure}

\caption{Average SSIM, with 99\% confidence intervals, for five reconstruction methods, using a generative adversarial network prior, which reconstructed one hundred $32\times 32$ test images (either CIFAR10 or CalTech101 downsampled to $32\times 32$ size). The average SSIM is presented as we either vary the compressive sensing ratio $\left(\frac{m}{n}\right)$ or the number of angles in the Radon transform. \textbf{Our A-CG-GAN method outperforms, or performs comparably to, the prior art methods in all scenarios}.}
\label{fig:compare}
\end{figure*}

We examine our proposed Algorithm~\ref{alg:CG-GAN} for two linear IPs in imaging, namely CS (where $\Psi\in\mathbb{R}^{m\times n}$ has entries sampled $\mathcal{N}(0, 1/m)$) and X-ray computed tomography (where $\Psi$ corresponds to a Radon transform at a number of uniformly spaced angles). Comparisons are made against four prior-art methods that similarly consider a GAN-prior in solving IPs, which we denote by Bora~\cite{ganbora}, Dhar~\cite{dhar2018deviations}, Shah~\cite{shah2018solving}, and Latorre~\cite{latorre2019GANADMM}. Note that Bora, Shah, and Latorre all find a solution to the IP within the range of a generative NN while Dhar considers a dual sparsity-GAN prior by finding a solution to the IP within sparse deviations from the range of a generative NN.

For data, we use $32\times 32$ CIFAR10~\cite{CIFAR10} images and CalTech101~\cite{CalTech101} images downsampled to size $64\times 64$ (and $32\times 32$). With the CIFAR10 and $64\times 64$ CalTech101 datasets, we set aside 100 images to serve as the testing data and we train a DCGAN~\cite{radford2015DCGAN} on the remaining images, which are augmented using rotation and reflection. We use $\overline{G}:\mathbb{R}^d\to\mathbb{R}^n$, with $d = 100$, to denote the DCGAN generator.

For training $\overline{G}$, which has a $\tanh$ output activation function, we scale and shift the training images onto the range $[-1,1]$ by replacing training image $\bm{I}$ by $\frac{2}{I_{\max}}\bm{I} - \bm{1}$ where $I_{\max}$ is the max pixel value of $\bm{I}$. For testing -- i.e., solving the IP to (\ref{eqn:linear_msrmt}) -- we scale the testing images onto the range $[0,1]$ by replacing test image $\bm{I}$ by $\bm{I}/I_{\max}.$ A measurement matrix $\Psi$ is applied to each test image, after which white noise is added, producing noisy measurements, $\bm{y}$, at a specified signal-to-noise-ratio (SNR). Additionally, for testing, we scale and shift the outputs from $\overline{G}$ onto the range $[0,1]$ by considering the generator $\widetilde{G}(\bm{x}) = \frac{\overline{G}(\bm{x})+\bm{1}}{2}$. Note that scaling to the range $[0,1]$ for testing is conducted so that we can use the structural similarity index measure (SSIM), which requires input images with positive pixel values, to assess the reconstruction performance.

For our A-CG-GAN algorithm, we will take the generative NN $G(\bm{x}) = \widetilde{\Phi} \widetilde{G}(\bm{x})$, where $\widetilde{\Phi}\in\mathbb{R}^{n\times n}$ is a sparsity change-of-basis matrix. We apply the change-of-basis matrix since we desire for the generative NN to produce sparse scale variables rather than the image directly. Furthermore, we set $\bm{\mu}_u = \bm{0}$, $\Sigma_u = I$, $R_x \equiv 0$, $\tau = 10^{-6}$, $K = 1000$, and $J_x = J = 10$ for all testing of A-CG-GAN. Instead, for each comparison method~\cite{ganbora, dhar2018deviations, shah2018solving, latorre2019GANADMM}, we use $G(\bm{x}) = \widetilde{G}(\bm{x})$ as the generative NN prior and a grid search is performed to find the best hyperparameters for each of the four comparison methods. Finally, for each comparison method, random restarts were implemented as specified in each work respectively~\cite{ganbora, shah2018solving, dhar2018deviations, latorre2019GANADMM}. We remark that using the same underlying image-generating NN provides fairness in the test IP comparisons since we guarantee the generative NN used in our method and each comparison method has the same training and quality.

Shown in Fig.~\ref{fig:compare} is the average SSIM -- larger values denote a reconstruction more closely matching the original image -- over the reconstructions of one hundred $32\times 32$ images from A-CG-GAN and each of the four comparison methods. Each plot in Fig.~\ref{fig:compare} gives both the average SSIM and 99\% confidence interval -- visualized by the background shading -- as the dimension of $\bm{y}$ is varied in both CS and CT problems. The underlying DCGAN generator, $\overline{G}$, used for each of these IPs has been trained on 50000 CIFAR10 images. Through a grid search, we set the hyperparameters of A-CG-GAN to be $(\lambda, \mu, \rho) = (0.1, 10^{-3}, 1)$ and $(\lambda, \mu, \rho) = (1, 10^{-2}, 1)$ for CT reconstructions from 60dB and 40dB, respectively. Furthermore, we set $(\lambda, \mu, \rho) = (10^{-6}, 10^{-6}, 10^{-4})$ as the hyperparameters of A-CG-GAN for CS reconstructions. Additionally, we use $\widetilde{\Phi} = \Phi = $ discrete cosine transformation and $\widetilde{\Phi} = \Phi = I$ for CS and CT problems, respectively.

In Figs.~\ref{fig:compare_CIFAR10_ratio_60}, \ref{fig:compare_CIFAR10_angles_60} and Figs.~\ref{fig:compare_CIFAR10_ratio_40}, \ref{fig:compare_CIFAR10_angles_40}, CIFAR10 images are reconstructed from CS and CT measurements with an SNR of 60dB and 40dB, respectively. Instead, in Figs.~\ref{fig:compare_CalTech101_ratio_60}, \ref{fig:compare_CalTech101_angles_60}, \ref{fig:compare_CalTech101_ratio_40}, and \ref{fig:compare_CalTech101_angles_40}, we show the generalizability of our method by reconstructing CalTech101 images, that have been downsampled to size $32\times 32$, while using the same CIFAR10 trained $\overline{G}$.

\begin{figure*}[!t]
\centering
\begin{subfigure}{0.24\textwidth}
    \centering
    \includegraphics[width=\textwidth]{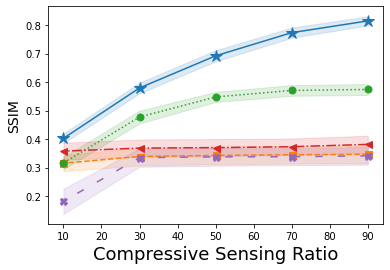}
    \caption{CalTech101 estimates from CS with 60dB SNR.}
    \label{fig:compare_CalTech101_64_ratio_60}
\end{subfigure}
\begin{subfigure}{0.24\textwidth}
    \centering
    \includegraphics[width=\textwidth]{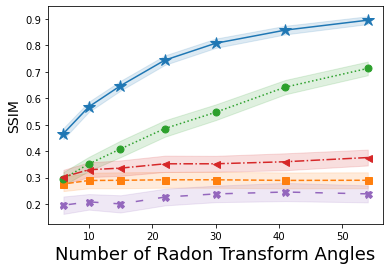}
    \caption{CalTech101 estimates from CT with 60dB SNR.}
    \label{fig:compare_CalTech101_64_angles_60}
\end{subfigure}
\begin{subfigure}{0.24\textwidth}
    \centering
    \includegraphics[width=\textwidth]{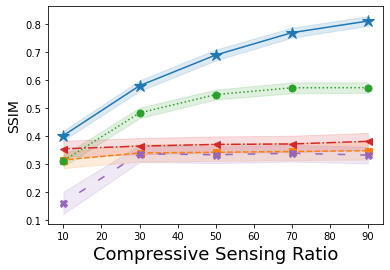}
    \caption{CalTech101 estimates from CS with 40dB SNR.}
    \label{fig:compare_CalTech101_64_ratio_40}
\end{subfigure}
\begin{subfigure}{0.24\textwidth}
    \centering
    \includegraphics[width=\textwidth]{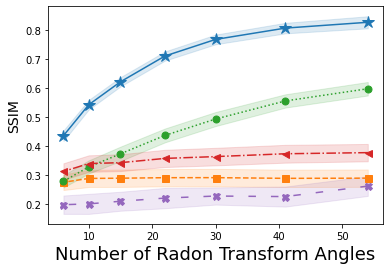}
    \caption{CalTech101 estimates from CT with 40dB SNR.}
    \label{fig:compare_CalTech101_64_angles_40}
\end{subfigure}
\begin{subfigure}{0.48\textwidth}
    \centering
    \includegraphics[width = \textwidth]{Images32Size/legend.PNG}
\end{subfigure}

\caption{Average SSIM, with 99\% confidence intervals, for five iterative reconstruction methods, using a generative adversarial network prior, which reconstructed one hundred $64\times 64$ test CalTech101 images. The average SSIM is presented as we either vary the compressive sensing ratio $\left(\frac{m}{n}\right)$ or the number of angles in the Radon transform. \textbf{Our A-CG-GAN method outperforms the compared prior art methods in all scenarios}.}
\label{fig:compare_64}
\end{figure*}

We observe, from Fig.~\ref{fig:compare}, that A-CG-GAN outperforms, or performs comparably to, each of the four prior art methods in all scenarios. In particular, we highlight that the performance of A-CG-GAN does not saturate as the dimension of $\bm{y}$ increases -- i.e., the sampling from the forward measurement mapping increases -- while the performance of the comparisons in general does. Additionally, A-CG-GAN is able to generalize well to image distributions outside of the GAN training distribution as shown in Figs.~\ref{fig:compare_CalTech101_ratio_60}, \ref{fig:compare_CalTech101_angles_60}, \ref{fig:compare_CalTech101_ratio_40}, and \ref{fig:compare_CalTech101_angles_40}. 

Next, shown in Fig.~\ref{fig:compare_64} is the average SSIM over the reconstructions of one hundred $64\times 64$ images from A-CG-GAN and each of the four comparison methods. Each plot in Fig.~\ref{fig:compare_64} gives both the average SSIM and 99\% confidence interval -- visualized by the background shading -- as the amount of sampling from the forward measurement mapping -- i.e., the dimension of $\bm{y}$ -- is varied in both CS and CT problems. The underlying DCGAN generator, $\overline{G}$, used for each of these IPs has been trained on a minimal dataset of roughly 9000 CalTech101 images of size $64\times 64$. Empirically, we choose $(\lambda, \mu, \rho) = 0.25\times (10^{-6}, 10^{-6}, 10^{-4})$ in A-CG-GAN for CS reconstructions, $(\lambda, \mu, \rho) = (10, 0.1, 10)$ for 40dB CT reconstructions, and otherwise use the same setup as in the $32\times 32$ image reconstruction cases.

We observe from Fig.~\ref{fig:compare_64}, that A-CG-GAN outperforms each of the four prior art methods in all scenarios. We highlight, in particular, that while a sub-optimal generator, due to a small training dataset, severely diminished the performance of each of the four comparison methods, A-CG-GAN still provides relatively high-quality reconstructed images as measured by SSIM. This is perhaps anticipated given the dual prior basis our method implements, which alleviates a complete dependence on the generative NN prior as in Bora, Shah, and Latorre~\cite{ganbora, shah2018solving, latorre2019GANADMM}. Additionally, we again observe, from Fig.~\ref{fig:compare_64}, that A-CG-GAN eliminates the performance saturation with increasing sampling from the forward measurement mapping.

Lastly, we remark that Dhar~\cite{dhar2018deviations} is, in general, the best comparison method to ours as seen in Fig.~\ref{fig:compare} and Fig.~\ref{fig:compare_64}. This provides further credence to the benefit of using dual prior information in addition to the GAN prior as Dhar, similar to our dual CG-GAN prior, incorporates a dual sparsity-GAN prior.

\section{Conclusion and Future Work}

Utilizing a dual-structured prior that consists of the powerful CG class of densities and a generative NN, we developed a novel iterative estimate algorithm for linear IPs. Specifically, we proposed an algorithm, called A-CG-GAN, that finds an IP solution satisfying the CG structure of (\ref{eqn:CG}) where the scale variable portion of the CG prior is constrained onto the range of a generative NN. Hence, within the informative CG class of distributions, the generative NN captures a rich scale variable statistical representation, which provides a wide, but informative, IP regularization. 

We conducted a theoretical analysis on the convergence of A-CG-GAN and showed that, under mild assumptions on the generative NN, a linear rate of convergence is guaranteed up to a neighborhood about a constrained minimizer solution. Subsequently, thorough numerical validation of A-CG-GAN in tomographic imaging and CS IPs was presented. Across multiple datasets and a wide array of forward measurement mappings, we empirically demonstrated that A-CG-GAN outperforms competitive state-of-the-art techniques to IPs that use a GAN prior. Specifically, A-CG-GAN displays three key properties missing from previous techniques using GAN priors for IPs:
\begin{enumerate}
    \item Performance does not saturate with less undersampling from the forward measurement mapping
    \item Generalization to alternative dataset distributions other than the GAN training distributions is possible
    \item A sub-optimal generative NN, possibly due to a lack of training data or inadequate training, still provides relatively high-quality reconstructions
\end{enumerate}

Building upon the new foundations established in this paper, we illuminate numerous opportunities for future exploration including: analyzing the performance of A-CG-GAN on larger image reconstructions and on data corrupted by different noise models. Studying each of these objectives can provide greater insight into the practical, real-world applicability of our method. 

Another intriguing line of work would be to use a generative NN in A-CG-GAN that has been trained to produce scale variables directly rather than sparsifying an image-generating NN as is done in this paper. We could anticipate a possible improvement in reconstruction performance over the numerical results presented in this paper as the generative NN would directly capture the statistics of the scale variables rather than implicitly capturing these statistics from the overall images. 

A last point of future work is to empirically study A-CG-GAN, and each of the four comparisons methods, when the generative NN has been trained on a single category of images. In particular, analyzing image reconstruction performance when the test images are from versus distinct from the training category. The datasets employed in this paper contain images from multiple categories -- e.g., CIFAR10 has 10 categories and CalTech101 has 101 categories -- creating a significant statistical variability in the datasets, which can result in lower quality images from a generative NN. Instead, when a generative NN is trained on a single category, it is often able to produce higher quality images from this single category. Accordingly, we conjecture that A-CG-GAN would handle the reconstructions of test images distinct from the training category significantly better than the four comparison methods, in much of the same manner that we observed in the $64\times 64$ CalTech101 reconstructions. Furthermore, we conjecture that A-CG-GAN (with $\bm{\mu}_u = \bm{1}$) may have minimal improvement, or be matched by, the four comparison methods when test images are sampled from the training category.

\section{Appendix}

In this appendix, we provide mathematical proof details for the main theoretical results of A-CG-GAN, presented in Section~\ref{sec:core theory}, by extending the proof ideas in~\cite{latorre2019GANADMM} from ADMM with a single GAN-prior constraint to our case of ADMM with dual-structured CG-GAN-prior constraints.
\subsection{Preliminary Tools}\label{apndx:preliminary tools}
\begin{lemma}[\hspace*{-3px}~\cite{latorre2019GANADMM}] \label{lemma:rewrite proximal operator}
    Let $g:\mathbb{R}^\ell\to\mathbb{R}$ be a differentiable function and $r:\mathbb{R}^\ell\to\mathbb{R}$ a convex possibly non-smooth function, then
    \begin{align*}
        &\textnormal{prox}_{\alpha r}(\bm{w} - \alpha \nabla_{\bm{w}} g(\bm{w}))  \\
        &\s= \scalebox{1.0}{$\arg\min_{\bm{t}}$}\,\, \langle \bm{t} - \bm{w}, \nabla_{\bm{w}} g(\bm{w}) \rangle + (2\alpha)^{-1} \norm{\bm{t}-\bm{w}}_2^2 + r(\bm{t}).
    \end{align*}
\end{lemma}

\begin{lemma}[\hspace*{-3px}~\cite{boyd2004convex}]\label{lemma:lipschitz cts grad implies inequality}
    Let $g:\mathbb{R}^\ell\to\mathbb{R}$ be differentiable and have $L_g$-Lipschitz continuous gradient. For any $\bm{w}, \widetilde{\bm{w}}\in\mathbb{R}^\ell$
    \begin{align}
        g(\bm{w}) \leq g(\widetilde{\bm{w}}) + \langle \bm{w} - \widetilde{\bm{w}}, \nabla_{\bm{w}} g(\widetilde{\bm{w}}) \rangle + \frac{L_g}{2}\norm{\bm{w} - \widetilde{\bm{w}}}_2^2. \label{eqn:smoothness of g}
    \end{align}
\end{lemma}

\begin{lemma}\label{lemma:ISTA step bound}
    Let $g:\mathbb{R}^\ell\to\mathbb{R}$ be a differentiable function satisfying (\ref{eqn:smoothness of g}) for some positive constant $L_g$, some $\widetilde{\bm{w}}\in\mathbb{R}^{\ell}$, and any $\bm{w}\in\mathbb{R}^{\ell}$. Let $r:\mathbb{R}^\ell\to\mathbb{R}$ be a convex and possibly non-smooth function. Let $\alpha\leq L_g^{-1}$. If $\widehat{\bm{w}} = \textnormal{prox}_{\alpha r}(\widetilde{\bm{w}} - \alpha \nabla_{\bm{w}} g(\widetilde{\bm{w}}))$, then for any $\theta\in [0,1]$ and $\bm{w}^*\in\mathbb{R}^\ell$
    \begin{align*}
        g(\widehat{\bm{w}}) + r(\widehat{\bm{w}}) &\leq g(\widetilde{\bm{w}}) + \theta\langle \bm{w}^*-\widetilde{\bm{w}}, \nabla_{\bm{w}} g(\widetilde{\bm{w}})\rangle \\
        &\s+ \frac{\theta^2}{2\alpha}  \norm{\bm{w}^*-\widetilde{\bm{w}}}_2^2 + \theta r(\bm{w}^*)+(1-\theta)r(\widetilde{\bm{w}}).
    \end{align*}
\end{lemma}
\begin{proof}
    By (\ref{eqn:smoothness of g}), $\alpha\leq L_g^{-1}$, and Lemma~\ref{lemma:rewrite proximal operator}
    \begin{align*}
        &g(\widehat{\bm{w}}) + r(\widehat{\bm{w}}) \\
        &\leq g(\widetilde{\bm{w}}) + \langle \widehat{\bm{w}} -\widetilde{\bm{w}}, \nabla_{\bm{w}} g(\widetilde{\bm{w}})\rangle + \frac{1}{2\alpha}  \norm{\widehat{\bm{w}}-\widetilde{\bm{w}}}_2^2 + r(\widehat{\bm{w}}) \\
        &\resizebox{\columnwidth}{!}{${\displaystyle= g(\widetilde{\bm{w}}) + \min_{\bm{t}}\left\{\langle \bm{t} -\widetilde{\bm{w}}, \nabla_{\bm{w}} g(\widetilde{\bm{w}})\rangle + \frac{1}{2\alpha}  \norm{\bm{t}-\widetilde{\bm{w}}}_2^2 + r(\bm{t})\right\}.}$}
    \end{align*}
    Bounding the final line by setting $\bm{t} = \theta\bm{w}^* + (1-\bm{\theta})\widetilde{\bm{w}}$ and then using convexity of $r$ produces the desired result.
\end{proof}

\begin{lemma}\label{lemma:lipschitz cts L}
   The function $\overline{\mathcal{L}}_{\rho}$ has Lipschitz continuous gradient with respect to $\bm{x}$, $\bm{u}$, and $\bm{c}$ separately.
\end{lemma}
\begin{proof}
    Observe
    \begin{align}
        & \begin{bmatrix}
            \nabla_{\bm{z}} \overline{\mathcal{L}}_{\rho}(\bm{v}, \pmb{\phi})^T &
            \nabla_{\bm{u}} \overline{\mathcal{L}}_{\rho}(\bm{v}, \pmb{\phi})^T &
            \nabla_{\bm{c}} \overline{\mathcal{L}}_{\rho}(\bm{v}, \pmb{\phi})^T
        \end{bmatrix}^T = \nonumber\\
        &\s\s\s \begin{bmatrix}
            \bm{\varphi}_{1} + \rho (\bm{z} - G(\bm{x})) - \bm{u}\odot \bm{\varphi}_2 + \rho \bm{u}\odot (\bm{z}\odot\bm{u} - \bm{c}) \\
           -\bm{z}\odot \bm{\varphi}_2 +\rho \bm{z}\odot(\bm{z}\odot\bm{u}-\bm{c}) \\
           \nabla_{\bm{c}} f_{\bm{y}}(\bm{c}) + \bm{\varphi}_2 + \rho(\bm{c} - \bm{z}\odot\bm{u})
        \end{bmatrix}, \label{eqn:L bar grad}
    \end{align}
    which implies 
    $\nabla_{\bm{z}}^2 \overline{\mathcal{L}}_{\rho}(\bm{v}, \pmb{\phi}) = \rho (D\{\bm{u}\}^2 + I)$, $\nabla_{\bm{u}}^2 \overline{\mathcal{L}}_{\rho}(\bm{v}, \pmb{\phi}) =  \rho D\{\bm{z}\}^2$, and $\nabla_{\bm{c}}^2\overline{\mathcal{L}}_{\rho}(\bm{v}, \pmb{\phi})  = f_{\bm{y}}(\bm{c}) + \rho I$.
    Therefore, $\norm{\nabla_{\bm{z}}^2 \overline{\mathcal{L}}_{\rho}(\bm{v}, \pmb{\phi}))}_2 \leq \rho (1+u_{\infty}^2)$, $\norm{\nabla_{\bm{u}}^2 \overline{\mathcal{L}}_{\rho}(\bm{v}, \pmb{\phi})}_2 \leq \rho z_{\infty}^2$, and $\norm{\nabla_{\bm{c}}^2 \overline{\mathcal{L}}_{\rho}(\bm{v}, \pmb{\phi})}_2 \leq L_f + \rho$
    since $f_{\bm{y}}$ has $L_f$-Lipschitz continuous gradient.
\end{proof}

\begin{lemma}\label{lemma:bound ||zeta_k+1 - zeta_k||}
    For any $j\in\{1, 2, 3, 4\}$ it holds that $\delta_{j,k}^2\leq 4\Delta_{j,k}^2.$
\end{lemma}
\begin{proof}
    Using the triangle inequality observe
\begin{align*}
    \delta_{j,k} &\leq \norm{\bm{\zeta}_j^{*}-\bm{\zeta}_j^{(k+1)}}_2^2 + \norm{\bm{\zeta}_j^{*}-\bm{\zeta}_j^{(k)}}_2^2 \leq 2\Delta_{j,k}. \qedhere
\end{align*}
\end{proof}

\begin{lemma}\label{lemma:dual variable norm bound}
    For any $j\in\{1, 2\}$ it holds that $\norm{\bm{\varphi}_{j,k}}_2\leq 4\sigma_0$.
\end{lemma}
\begin{proof}
    Since $\bm{\varphi}_{j, 0} = \bm{0}$ and $\bm{\varphi}_{j,k} = \bm{\varphi}_{j,k-1} + \sigma_{k}\bm{\xi}_{j,k}$, then by the triangle inequality and (\ref{eqn:sigma_k for theory})
        \begin{align*}
        &\norm{\bm{\varphi}_{j,k}}_2 = \left\lVert\sum_{i=1}^k \sigma_{i}\bm{\xi}_{j,i}\right\rVert_2 \leq \sum_{i=1}^k \sigma_i\norm{\bm{\xi}_{j,i}}_2 \leq \sum_{i=1}^k \frac{\sigma_{0}}{i\ln^2(i+1)}.
    \end{align*}
    Finally, note that $\sum_{i=1}^k \frac{1}{i\ln^2(i+1)}\leq \sum_{i=1}^\infty \frac{1}{i\ln^2(i+1)} \leq 4$.
\end{proof}

\subsection{Dual-Variable Update Bound}\label{apndx:proof of corollary:dual-variable update bound}
\begin{lemma}\label{lemma:latent variable bound}
    For any $k\in\mathbb{N}$ there exists positive constants $\{\gamma_i\}_{i = 1}^4$ such that $\sigma_{k+1}\norm{\bm{\xi}_{k+1}}_2^2 \leq 2\sigma_0\sum_{j=1}^4 \gamma_i \Delta_{i,k}^2.$
\end{lemma}
\begin{proof}
    Using the feasibility condition $\bm{\xi}^* = \bm{0}$, the triangle inequality, Assumption~\ref{assume:G}, and $(a+b)^2\leq 2(a^2+b^2)$ observe
        \begin{align*}
    \norm{\bm{\xi}_{1,k+1}}_2^2 
    &= \norm{\bm{z}_{k+1} - \bm{z}^* + G(\bm{x}^*) - G(\bm{x}_{k+1})}_2^2 \\
    &\leq 2\Delta_{2,k}^2 + 2\tau_G^2\Delta_{1,k}^2
    \end{align*}
    and
    \begin{align*}
        \norm{\bm{\xi}_{2,k+1}}_2^2 
        &= \norm{\bm{c}_{k+1} - \bm{c}^* + \bm{z}^*\odot\bm{u}^* - \bm{z}_{k+1}\odot\bm{u}_{k+1}}_2^2 \\
        &\leq 2\Delta_{4,k}^2 + (2z_{\infty})^2 \Delta_{3,k}^2 + (2u_\infty)^2 \Delta_{2,k}^2.
    \end{align*}
    Combining the above two inequalities with $\sigma_{k+1} \leq \sigma_0$ and $\norm{\bm{\xi}_{k+1}}_2^2 = \norm{\bm{\xi}_{1,k+1}}_2^2 + \norm{\bm{\xi}_{2,k+1}}_2^2$ produces the desired bound for $\gamma_1 = \tau_G^2$, $\gamma_2 = 1+2u_\infty^2$, $\gamma_3 = 2z_{\infty}^2$, and $\gamma_4 = 1$.
\end{proof}

\begin{corollary}\label{corollary:dual-variable update bound}
    For every $k\in\mathbb{N}$
    \begin{align*}
        \overline{\mathcal{L}}_{k+1} \leq \overline{\mathcal{L}}_{\rho}(\bm{v}_{1:4}^{(k+1)}, \pmb{\phi}_k) + 2\sigma_0\sum_{j=1}^4 \gamma_i \Delta_{i,k}^2.
    \end{align*}
\end{corollary}
\begin{proof}
Using that $\pmb{\phi}_{k+1} = \pmb{\phi}_k + \sigma_{k+1} \bm{\xi}_{k+1}$ with (\ref{eqn:augmented Lagrangian generalized adjusted}) and then applying Lemma~\ref{lemma:latent variable bound} produces the desired bound.
\end{proof}

\subsection{Proof of Proposition~\ref{prop:w update bound}}\label{apndx:proof of prop:w update bound}

Define $\widehat{\bm{c}}_k = \text{prox}_{\alpha_c R_4}(\bm{c}_k - \alpha_4 \nabla_{\bm{c}} \overline{\mathcal{L}}_{\rho}(\bm{v}_{1:3}^{(k+1)}, \bm{c}_k, \pmb{\phi}_k))$ to be the ISTA/PGD update of $\overline{\mathcal{L}}_{\rho}$ w.r.t $\bm{c}$ at $\bm{c}_k$. Since $\bm{c}_{k+1} = \arg\min_{\bm{c}}\overline{\mathcal{L}}_{\rho}(\bm{v}_{1:3}^{(k+1)}, \bm{c}, \pmb{\phi}_k) + R_4(\bm{c})$ then
\begin{align*}
    \overline{\mathcal{L}}_{\rho}(\bm{v}_{1:4}^{(k+1)}, \pmb{\phi}_k) + R_4(\bm{c}_{k+1}) \leq \overline{\mathcal{L}}_{\rho}(\bm{v}_{1:3}^{(k+1)}, \widehat{\bm{c}}_k, \pmb{\phi}_k) + R_4(\widehat{\bm{c}}_k).
\end{align*}
Combining this inequality with Lemmas~\ref{lemma:lipschitz cts L}, \ref{lemma:lipschitz cts grad implies inequality}, and \ref{lemma:ISTA step bound} gives
\begin{align}
        &\overline{\mathcal{L}}_{\rho}(\bm{v}_{1:4}^{(k+1)}, \pmb{\phi}_k) + R_4(\bm{c}_{k+1}) \nonumber \\
        &\leq \overline{\mathcal{L}}_{\rho}(\bm{v}_{1:3}^{(k+1)}, \bm{c}_k, \pmb{\phi}_k) + \theta R_4(\bm{c}^*) + (1-\theta)R_4(\bm{c}_k) \nonumber \\
    &\s + \frac{\theta^2}{2\alpha_c} \Delta_{4,k}^2 + \theta\langle \bm{c}^*-\bm{c}_k, \nabla_{\bm{c}} \overline{\mathcal{L}}_{\rho}(\bm{v}_{1:3}^{(k+1)}, \bm{c}_k, \pmb{\phi}_k) \rangle  \label{eqn:c update bound 1}
    \end{align}
where, in Lemma~\ref{lemma:ISTA step bound}, we take $g(\bm{c}) = \overline{\mathcal{L}}_{\rho}(\bm{v}_{1:3}^{(k+1)}, \bm{c}, \pmb{\phi}_k)$ (which has $(L_f+\rho)$-Lipschitz continuous gradient by Lemma~\ref{lemma:lipschitz cts L} and thus satisfies (\ref{eqn:smoothness of g}) by Lemma~\ref{lemma:lipschitz cts grad implies inequality}), $r = R_4$, $L_g = L_f+\rho$, $\widetilde{\bm{w}} = \bm{c}_k$, and $\bm{w}^* = \bm{c}^*.$

Next, by (\ref{eqn:L bar grad}), for any $\bm{x}, \widehat{\bm{x}}, \widetilde{\bm{x}} \in\mathbb{R}^d$ and any $\bm{z}, \widehat{\bm{z}}, \widetilde{\bm{z}}, \bm{u}, \widehat{\bm{u}}, \widetilde{\bm{u}}, \bm{c} \in\mathbb{R}^n$ it holds that
\begin{align*}
    &\nabla_{\bm{c}} \overline{\mathcal{L}}_{\rho}(\widehat{\bm{x}}, \bm{v}_{2:4}, \pmb{\phi}) -  \nabla_{\bm{c}} \overline{\mathcal{L}}_{\rho}(\widetilde{\bm{x}}, \bm{v}_{2:4}, \pmb{\phi}) = \bm{0} \\
    &\resizebox{\columnwidth}{!}{${\displaystyle \nabla_{\bm{c}} \overline{\mathcal{L}}_{\rho}(\bm{x}, \widehat{\bm{v}}_{2:3}, \bm{c}, \pmb{\phi}) - \nabla_{\bm{c}} \overline{\mathcal{L}}_{\rho}(\bm{x}, \widetilde{\bm{v}}_{2:3}, \bm{c}, \pmb{\phi}) = -\rho (\widehat{\bm{z}}\odot\widehat{\bm{u}} - \widetilde{\bm{z}}\odot\widetilde{\bm{u}}).}$}
\end{align*}
Thus
\begin{align}
    &\theta\langle \bm{c}^*-\bm{c}_k, \nabla_{\bm{c}} \overline{\mathcal{L}}_{\rho}(\bm{v}_{1:3}^{(k+1)}, \bm{c}_k, \pmb{\phi}_k) \rangle - \theta\langle \bm{c}^*-\bm{c}_k, \nabla_{\bm{c}} \overline{\mathcal{L}}_{k}\rangle \nonumber \\
    &\s = \theta\langle \bm{c}^*-\bm{c}_k, -\rho (\bm{z}_{k+1}\odot\bm{u}_{k+1} - \bm{z}_k\odot\bm{u}_k) \rangle \nonumber \\
    &\s = \theta\langle \bm{c}^*-\bm{c}_k, -\rho (\bm{z}_{k+1}-\bm{z}_k)\odot\bm{u}_{k+1} \rangle   \nonumber \\
    &\s\s + \theta\langle \bm{c}^*-\bm{c}_k, -\rho \bm{z}_k\odot (\bm{u}_{k+1}-\bm{u}_k) \rangle \nonumber \\
    &\s \leq \theta^2 \Delta_{4,k}^2 + \frac{(\rho u_{\infty})^2}{2} \delta_{2,k}^2 + \frac{(\rho z_{\infty})^2}{2} \delta_{3,k}^2 \label{eqn:c update bound 2}
\end{align}
where we use $2\langle \bm{a}, \bm{b}\rangle \leq \norm{\bm{a}}_2^2 + \norm{\bm{b}}_2^2$ in the final inequality. Combining (\ref{eqn:c update bound 1}) and (\ref{eqn:c update bound 2}) produces the desired result. \qed

\subsection{Proof of Proposition~\ref{prop:u update bound}}\label{apndx:proof of prop:u update bound}

Let $\widehat{\bm{u}}_k = \text{prox}_{\alpha_u R_3}\left(\bm{u}_k - \alpha_u \nabla_{\bm{u}}\overline{\mathcal{L}}_{\rho}(\bm{v}_{1:2}^{(k+1)}, \bm{u}_k, \bm{c}_k, \pmb{\phi}_k)\right)$ be the ISTA/PGD update of $\overline{\mathcal{L}}_{\rho}$ w.r.t $\bm{u}$ at $\bm{u}_k$. Since $\bm{u}_{k+1} = \arg\min_{\bm{u}} \overline{\mathcal{L}}_{\rho}(\bm{v}_{1:2}^{(k+1)}, \bm{u}, \bm{c}_k, \pmb{\phi}_k)$ then
\begin{multline}
    \overline{\mathcal{L}}_{\rho}(\bm{v}_{1:3}^{(k+1)}, \bm{c}_k, \pmb{\phi}_k) + R_3(\bm{u}_{k+1})  \\
     \leq  \overline{\mathcal{L}}_{\rho}(\bm{v}_{1:2}^{(k+1)}, \widehat{\bm{u}}_k , \bm{c}_k, \pmb{\phi}_k) + R_3(\widehat{\bm{u}}_k). \label{eqn:u update bound 0}
\end{multline}
Define $g(\bm{u}) = \overline{\mathcal{L}}_{\rho}(\bm{v}_{1:2}^{(k+1)}, \bm{u}, \bm{c}_k, \pmb{\phi}_k)$, which has $(\rho z_{\infty}^2)$-Lipschitz continuous gradient by Lemma~\ref{lemma:lipschitz cts L}. Hence, $g(\bm{u})$ satisfies (\ref{eqn:smoothness of g}) by Lemma~\ref{lemma:lipschitz cts grad implies inequality}. Combining (\ref{eqn:u update bound 0}) with Lemma~\ref{lemma:ISTA step bound} where $g = g(\bm{u}), r = R_3, L_g = \rho z_{\infty}^2, \widetilde{\bm{w}} = \bm{u}_k,$ and $\bm{w}^* = \bm{u}^*$ gives
    \begin{align}
        & \overline{\mathcal{L}}_{\rho}(\bm{v}_{1:3}^{(k+1)}, \bm{c}_k, \pmb{\phi}_k) + R_3(\bm{u}_{k+1}) \nonumber \\
        &\leq \overline{\mathcal{L}}_{\rho}(\bm{v}_{1:2}^{(k+1)}, \bm{v}_{3:4}^{(k)}, \pmb{\phi}_k)  + \theta R_3(\bm{u}^*) + (1-\theta)R_3(\bm{u}_k) \nonumber \\
        &\s + \frac{\theta^2}{2\alpha_u} \Delta_{3,k}^2  + \theta\langle \bm{u}^* - \bm{u}_k, \nabla_{\bm{u}} \overline{\mathcal{L}}_{\rho}(\bm{v}_{1:2}^{(k+1)}, \bm{v}_{3:4}^{(k)}, \pmb{\phi}_k)\rangle. \label{eqn:u update bound 1}
    \end{align}

Next, by (\ref{eqn:L bar grad}), for any $\bm{x}, \widehat{\bm{x}}, \widetilde{\bm{x}} \in\mathbb{R}^d$ and any $\bm{z}, \widehat{\bm{z}}, \widetilde{\bm{z}}, \bm{u}, \bm{c}, \widehat{\bm{c}}, \widetilde{\bm{c}} \in\mathbb{R}^n$ it holds that
\begin{align*}
    \nabla_{\bm{u}} \overline{\mathcal{L}}_{\rho}(\widehat{\bm{x}}, \bm{v}_{2:4}, \pmb{\phi}) -  \nabla_{\bm{u}} \overline{\mathcal{L}}_{\rho}(\widetilde{\bm{x}}, \bm{v}_{2:4}, \pmb{\phi}) &= \bm{0}
\end{align*}
and
\begin{multline*}
    \nabla_{\bm{u}} \overline{\mathcal{L}}_{\rho}(\bm{x}, \widehat{\bm{z}}, \bm{v}_{3:4}, \pmb{\phi}) - \nabla_{\bm{u}} \overline{\mathcal{L}}_{\rho}(\bm{x}, \widetilde{\bm{z}}, \bm{v}_{3:4}, \pmb{\phi}) \\
     = (\rho\bm{u}\odot(\widehat{\bm{z}} + \widetilde{\bm{z}})-\bm{\varphi}_2-\rho\bm{c})\odot (\widehat{\bm{z}}-\widetilde{\bm{z}}).
\end{multline*}
Thus
\begin{align}
    &\theta\langle \bm{u}^* - \bm{u}_k, \nabla_{\bm{u}} \overline{\mathcal{L}}_{\rho}(\bm{v}_{1:2}^{(k+1)}, \bm{v}_{3:4}^{(k)}, \pmb{\phi}_k)\rangle - \theta\langle \bm{u}^* - \bm{u}_k, \nabla_{\bm{u}} \overline{\mathcal{L}}_k\rangle \nonumber \\
    &\resizebox{\columnwidth}{!}{${\displaystyle = \theta\langle \bm{u}^* - \bm{u}_k, (\rho\bm{u}_{k}\odot(\bm{z}_{k+1} + \bm{z}_k)-\bm{\varphi}_{2,k}-\rho\bm{c}_k)\odot (\bm{z}_{k+1}-\bm{z}_k)\rangle }$} \nonumber \\ 
    & \leq \frac{\theta^2}{2}\Delta_{3,k}^2 + \frac{(2\rho u_{\infty}z_\infty + 4\sigma_0 + \rho c_\infty)^2}{2}\delta_{2,k}^2 \label{eqn:u update bound 2}
\end{align}
where in the final inequality we use $2\langle a, b\rangle \leq \norm{a}_2^2 + \norm{b}_2^2$ and Lemma~\ref{lemma:dual variable norm bound}. Combining (\ref{eqn:u update bound 1}) and (\ref{eqn:u update bound 2}) produces the desired result with $\gamma = (2\rho u_{\infty}z_\infty + 4\sigma_0 + \rho c_\infty)^2.$ \qed

\subsection{Proof of Proposition~\ref{prop:z update bound}}\label{apndx:proof of prop:z update bound}

Define $g(\bm{z}) \coloneqq \overline{\mathcal{L}}_{\rho}(\bm{x}_{k+1}, \bm{z}, \bm{v}_{3:4}^{(k)}, \pmb{\phi}_k)$ and let $\widehat{\bm{z}}_k = \text{prox}_{\alpha_z R_2}\left(\bm{z}_k - \alpha_z \nabla_{\bm{z}}g(\bm{z}_k)\right)$ be a single ISTA step on $g$ w.r.t $\bm{z}$ from $\bm{z}_k.$ Since $\bm{z}_{k+1}$ is the output of multiple FISTA/ISTA steps on $g+R_2$ w.r.t $\bm{z}$ from $\bm{z}_k$, which decrease the cost function at least as much as a single ISTA step, then
\begin{multline}
    \overline{\mathcal{L}}_{\rho}(\bm{v}_{1:2}^{(k+1)}, \bm{v}_{3:4}^{(k)},  \pmb{\phi}_k) + R_2(\bm{z}_{k+1})  \\
     \leq  \overline{\mathcal{L}}_{\rho}(\bm{x}_{k+1},\widehat{\bm{z}}_k, \bm{v}_{3:4}^{(k)}, \pmb{\phi}_k) + R_2(\widehat{\bm{z}}_k). \label{eqn:z update bound 0}
\end{multline}
Note that $g$ has $(\rho(1+u_{\infty}^2))$-Lipschitz continuous gradient by Lemma~\ref{lemma:lipschitz cts L}. Hence, $g(\bm{z})$ satisfies (\ref{eqn:smoothness of g}) by Lemma~\ref{lemma:lipschitz cts grad implies inequality}. Combining (\ref{eqn:z update bound 0}) with Lemma~\ref{lemma:ISTA step bound} where $g = g(\bm{z}), r = R_2, L_g = \rho(1+u_{\infty}^2), \widetilde{\bm{w}} = \bm{z}_k,$ and $\bm{w}^* = \bm{z}^*$ gives
\begin{align}
    & \overline{\mathcal{L}}_{\rho}(\bm{v}_{1:2}^{(k+1)}, \bm{v}_{3:4}^{(k)},  \pmb{\phi}_k) + R_2(\bm{z}_{k+1}) \nonumber \\
    &\leq \overline{\mathcal{L}}_{\rho}(\bm{x}_{k+1},\bm{v}_{2:4}^{(k)}, \pmb{\phi}_k) + \theta\langle \bm{z}^* - \bm{z}_k, \nabla_{\bm{z}} \overline{\mathcal{L}}_{\rho}(\bm{x}_{k+1},\bm{v}_{2:4}^{(k)}, \pmb{\phi}_k)\rangle  \nonumber \\
    &\hspace{.45cm}  + \frac{\theta^2}{2\alpha_z} \Delta_{2,k}^2 + \theta R_z(\bm{z}^*) + (1-\theta)R_z(\bm{z}_k). \label{eqn:z update bound 1}
\end{align}

Next, by (\ref{eqn:L bar grad}), for any $\bm{x}, \widehat{\bm{x}}, \widetilde{\bm{x}} \in\mathbb{R}^d$ and any $\bm{z}, \bm{u}, \bm{c}, \in\mathbb{R}^n$ it holds that
    \begin{align*}
        \nabla_{\bm{z}} \overline{\mathcal{L}}_{\rho}(\widehat{\bm{x}}, \bm{v}_{2:4},\pmb{\phi}) - \nabla_{\bm{z}} \overline{\mathcal{L}}_{\rho}(\widetilde{\bm{x}}, \bm{v}_{2:4},\pmb{\phi}) = -\rho (G(\widehat{\bm{x}}) - G(\widetilde{\bm{x}})).
    \end{align*}
Thus, using $2\langle \bm{a}, \bm{b}\rangle \leq \norm{\bm{a}}_2^2 + \norm{\bm{b}}_2^2$ and Assumption~\ref{assume:G}
    \begin{align*}
        &\theta\langle \bm{z}^*- \bm{z}_k, \nabla_{\bm{z}} \overline{\mathcal{L}}_{\rho}(\bm{x}_{k+1},\bm{v}_{2:4}^{(k)}, \pmb{\phi}_k)\rangle - \theta\langle \bm{z}^* - \bm{z}_k, \nabla_{\bm{z}} \overline{\mathcal{L}}_k\rangle \\
        & = \theta\langle \bm{z}^*- \bm{z}_k, -\rho(G(\bm{x}_{k+1}) - G(\bm{x}_k))\rangle  \\
        & \leq \frac{\theta^2}{2}\Delta_{2,k}^2 + \frac{\rho^2\tau_G^2}{2}\norm{\bm{x}_{k+1} -\bm{x}_k}_2^2.
    \end{align*}
    Combining this inequality with (\ref{eqn:z update bound 1}) produces the desired results. \qed

\subsection{Bound on \textit{x} Update (Proof of Proposition~\ref{prop:x update bound})}\label{apndx:proof of prop:x update bound}

First, we provide two lemmas to simplify the proof of Proposition~\ref{prop:x update bound}. For these lemmas, define $\varrho_k:\mathbb{R}^d\to\mathbb{R}$ as
\begin{multline*}
    \varrho_k(\bm{x}) \coloneqq ||\bm{z}_k-G(\bm{x})||_2^2 - ||\bm{\xi}_{1,k}||_2^2 \\
    + 2\langle (\nabla_{\bm{x}} G(\bm{x}_k))^T(\bm{x}-\bm{x}_k),  \bm{\xi}_{1,k})\rangle.
\end{multline*}
\begin{lemma}\label{lemma:x update bound 1}
For every $k\in\mathbb{N}$ and any $\bm{x}\in\mathbb{R}^d$
\begin{align*}
    \varrho_k(\bm{x}) \leq (\tau_G^2 + L_G \norm{\bm{\xi}_{1,k}}_2)\norm{\bm{x}-\bm{x}_k}_2^2.
\end{align*}
\end{lemma}
\begin{proof}
First note that for any $\bm{t}, \widehat{\bm{t}}, \widetilde{\bm{t}}\in\mathbb{R}^n$ it holds that
    \begin{align*}
        \norm{\bm{t}-\widehat{\bm{t}}}_2^2 - \norm{\bm{t}-\widetilde{\bm{t}}}_2^2 + 2\langle \widehat{\bm{t}}-\widetilde{\bm{t}},\bm{t}-\widetilde{\bm{t}}\rangle = \norm{\widehat{\bm{t}}-\widetilde{\bm{t}}}_2^2.
    \end{align*}
Therefore
\begin{align*}
    \varrho_k(\bm{x})  & = \norm{G(\bm{x}) - G(\bm{x}_k)}_2^2 \\
    &\s\s + 2\langle G(\bm{x}_{k})-G(\bm{x})+(\nabla_{\bm{x}} G(\bm{x}_k))^T(\bm{x}-\bm{x}_k),\bm{\xi}_{1,k}\rangle \\
    &\leq \tau_G^2 \norm{\bm{x}-\bm{x}_k}_2^2 + L_G\norm{\bm{\xi}_{1,k}}_2\norm{\bm{x}-\bm{x}_k}_2^2
\end{align*}
where we use the Cauchy-Schwarz inequality and Assumption~\ref{assume:G} to furnish the final inequality.
\end{proof}

\begin{lemma}\label{lemma:x update bound 2}
For every $k\in\mathbb{N}$ any $\bm{w}\in\mathbb{R}^d$ the inequality (\ref{eqn:smoothness of g}) holds for $g = g(\bm{x}) = \overline{\mathcal{L}}_{\rho}(\bm{x}, \bm{v}_{2:4}^{(k)}, \pmb{\phi}_k), \widetilde{\bm{w}} = \bm{x}_k$, and $L_g = L_G(4\sigma_0+\rho \norm{\bm{\xi}_{1,k}}_2)+\rho\tau_G^2$.
\end{lemma}
\begin{proof}
    First, note that
\begin{align*}
   \nabla_{\bm{x}} g(\bm{x}_k) & = -\nabla_{\bm{x}} G(\bm{x}_k)\left(\bm{\varphi}_{1,k} + \rho\bm{\xi}_{1,k}\right).
\end{align*}
Now, observe
\begin{align*}
    & g(\bm{x}) - g(\bm{x}_k) - \langle \bm{x}-\bm{x}_k, \nabla_{\bm{x}} g(\bm{x}_k)\rangle \\
    &\resizebox{\columnwidth}{!}{${\displaystyle = \langle \bm{\varphi}_{1,k}, G(\bm{x}_k) - G(\bm{x}) + (\nabla_{\bm{x}} G(\bm{x}_k))^T(\bm{x}-\bm{x}_k) \rangle + \frac{\rho}{2}\varrho_k(\bm{x}) }$} \\
    &\leq \frac{4\sigma_0 L_G}{2}\norm{\bm{x}-\bm{x}_k}_2^2 + \frac{\rho(\tau_G^2 + L_G\norm{\bm{\xi}_{1,k}}_2)}{2}\norm{\bm{x}-\bm{x}_k}_2^2
\end{align*} 
where we use the Cauchy-Schwarz inequality, Assumption~\ref{assume:G}, Lemma~\ref{lemma:dual variable norm bound}, and Lemma~\ref{lemma:x update bound 1} to produce the final inequality.
\end{proof}

We now prove the desired bound on the adjusted augmented Lagrangian over an update of $\bm{x}.$
\begin{proof}[Proof of Proposition~\ref{prop:x update bound}]
    Define $g(\bm{x}) = \overline{\mathcal{L}}_{\rho}(\bm{x}, \bm{v}_{2:4}^{(k)}, \pmb{\phi}_k)$ and let $\widehat{\bm{x}}_k = \text{prox}_{\alpha_1 R_1}\left(\bm{x}_k - \alpha_1 \nabla_{\bm{x}}g(\bm{x}_k)\right).$ Since $\bm{x}_{k+1}$ is the output of many GD/ISTA steps on $g+R_1$ w.r.t $\bm{x}$ from $\bm{x}_k$, which decrease the cost function at least as much as a single ISTA step, then
    \begin{multline}
        \overline{\mathcal{L}}_{\rho}(\bm{x}_{k+1}, \bm{v}_{2:4}^{(k)}, \pmb{\phi}_k) + R_1(\bm{x}_{k+1}) \\
        \leq  \overline{\mathcal{L}}_{\rho}(\widehat{\bm{x}}_k, \bm{v}_{2:4}^{(k)}, \pmb{\phi}_k) + R_1(\widehat{\bm{x}}_k). \label{eqn:x update bound 0}
    \end{multline}
    By Lemma~\ref{lemma:x update bound 2} and Lemma~\ref{lemma:ISTA step bound} with $g = g(\bm{x}), r = R_1, \widetilde{\bm{w}} = \bm{x}_k, L_g =  L_G(4\sigma_0+\rho \norm{\bm{\xi}_{1,k}}_2)+\rho\tau_G^2,$ and $\bm{w}^* = \bm{x}^*$ we have
    \begin{multline*}
        \overline{\mathcal{L}}_{\rho}(\widehat{\bm{x}}_k, \bm{v}_{2:4}^{(k)}, \pmb{\phi}_k) + R_1(\widehat{\bm{x}}_k) \\
        \leq  \overline{\mathcal{L}}_k + \theta\langle \bm{x}^*-\bm{x}_k, \nabla_{\bm{x}} \overline{\mathcal{L}}_k\rangle+ \frac{\theta^2}{\alpha_1}\Delta_{1,k}^2 \\
           + \theta R_1(\bm{x}^*) + (1-\theta) R_1(\bm{x}_k),
    \end{multline*}
    which combined with (\ref{eqn:x update bound 0}) produces the desired result.
\end{proof}

\subsection{Complete Iteration Bound (Proof of Proposition~\ref{prop:all variable update + full inner product bound})}\label{apndx:proof of prop:all variable update + full inner product bound}

In order to prove Proposition~\ref{prop:all variable update + full inner product bound} we first derive two simplifying lemmas.

\begin{lemma}\label{lemma:full inner product bound prelim 1}
   Let $\bm{q}_i\in\mathbb{R}^n$ for $i \in\{1,2\}$ and $\bm{q} = \begin{bmatrix}
       \bm{q}_1^T & \bm{q}_2^T
   \end{bmatrix}^T.$ For every $k\in\mathbb{N}$
      \begin{multline*}
    \left|\left\langle \bm{\xi}_k + (\nabla_{\bm{v}}\bm{\xi}_k)^T(\bm{v}^*-\bm{v}_k), \bm{q} \right\rangle \right| \\
    \leq \frac{L_G\norm{\bm{q}_1}_2}{2} \Delta_{1,k}^2 + \frac{\norm{\bm{q}_2}_2}{2}\left(\Delta_{2,k}^2 + \Delta_{3,k}^2\right).
   \end{multline*}
\end{lemma}
\begin{proof}
      First, note that
      \begin{align*}
          \resizebox{\columnwidth}{!}{${\displaystyle(\nabla_{\bm{v}}\bm{\xi}_k)^T =\begin{bmatrix}
           -(\nabla_{\bm{x}} G(\bm{x}_k))^T & I_{n\times n} & 0 & 0 \\
            0 & -D\{\bm{u}_k\} & -D\{\bm{z}_k\} & I_{n\times n}
            \end{bmatrix},}$} 
        \end{align*}
    which combined with the feasibility condition $\bm{\xi}^* = \bm{0}$ gives
    \begin{align*}
        &\bm{\xi}_k + (\nabla_{\bm{v}}\bm{\xi}_k)^T(\bm{v}^*-\bm{v}_k) \\
        &\s\s = \begin{bmatrix}
           \bm{z}^* - G(\bm{x}_k) -(\nabla_{\bm{x}} G(\bm{x}_k))^T(\bm{x}^*-\bm{x}_k) \\
           \bm{c}^* -\bm{z}^*\odot \bm{u}_k - \bm{z}_k\odot\bm{u}^* + \bm{z}_k\odot\bm{u}_k
        \end{bmatrix} \\
        &\s\s = \begin{bmatrix}
           G(\bm{x}^*) - G(\bm{x}_k) -(\nabla_{\bm{x}} G(\bm{x}_k))^T(\bm{x}^*-\bm{x}_k) \\
           (\bm{z}^*-\bm{z}_k)\odot(\bm{u}^*-\bm{u}_k)
        \end{bmatrix}.
    \end{align*} 
    Since $2ab\leq a^2+b^2$, note that
    \begin{align}
        2\norm{(\bm{z}^*-\bm{z}_k)\odot(\bm{u}^*-\bm{u}_k)}_2 \leq \Delta_{2,k}^2 + \Delta_{3,k}^2. \label{eqn:split (z*-zk)(u*-uk)}
    \end{align}
    Therefore, by the Cauchy-Schwarz and triangle inequalities
    \begin{multline*}
        \left|\left\langle \bm{\xi}_k + (\nabla_{\bm{v}}\bm{\xi}_k)^T(\bm{v}^*-\bm{v}_k), \bm{q} \right\rangle\right| \\
        \leq \norm{G(\bm{x}^*) - G(\bm{x}_k) -(\nabla_{\bm{x}} G(\bm{x}_k))^T(\bm{x}^*-\bm{x}_k)}_2\norm{\bm{q}_1}_2 \\
         + \norm{(\bm{z}^*-\bm{z}_k)\odot(\bm{u}^*-\bm{u}_k)}_2\norm{\bm{q}_2}_2.
    \end{multline*}
    Finally, applying Assumption~\ref{assume:G} and (\ref{eqn:split (z*-zk)(u*-uk)}) produces the desired result.
\end{proof}

\begin{lemma}\label{lemma:full inner product bound}
    Let $\psi_{j,k} = 4\sigma_0 + \rho \norm{\bm{\xi}_{j,k}}_2$ for $j\in\{1,2\}$ and $k\in\mathbb{N}$. For every $k\in\mathbb{N}$
    \begin{multline*}
        \overline{\mathcal{L}}_k - \overline{\mathcal{L}}_* + \langle \bm{v}^* - \bm{v}_{k}, \nabla_{\bm{v}} \overline{\mathcal{L}}_{k} \rangle \\
        \leq \frac{L_G \psi_{1,k}}{2}\Delta_{1,k}^2 + \frac{\psi_{2,k}}{2}\left(\Delta_{2,k}^2 + \Delta_{3,k}^2\right).
    \end{multline*}
\end{lemma}
\begin{proof}
    First, note that 
    \begin{multline}
    \nabla_{\bm{v}}\overline{\mathcal{L}}_{k} = \begin{bmatrix}
    \bm{0}^T & \bm{0}^T & \bm{0}^T & (\nabla_{\bm{c}} f_{\bm{y}}(\bm{c}_k))^T
    \end{bmatrix}^T \\
    + (\nabla_{\bm{v}}\bm{\xi}_k)\left(\pmb{\phi}_k + \rho \bm{\xi}_k\right). \label{eqn:gradient of line L}
    \end{multline}
    Hence
    \begin{align*}
        &\overline{\mathcal{L}}_k - \overline{\mathcal{L}}_* + \langle \bm{v}^* - \bm{v}_{k}, \nabla_{\bm{v}} \overline{\mathcal{L}}_{k} \rangle \\
        & = f_{\bm{y}}(\bm{c}_k) - f_{\bm{y}}(\bm{c}^*) + \langle \bm{c}^*-\bm{c}_k, \nabla_{\bm{c}} f_{\bm{y}}(\bm{c}_k)\rangle \\
        &\s\s + \langle \pmb{\phi}_k, \bm{\xi}_k\rangle + \frac{\rho}{2}\norm{\bm{\xi}_k}_2^2 + \left\langle \bm{v}^* - \bm{v}_k, (\nabla_{\bm{v}}\bm{\xi}_k)\left(\pmb{\phi}_k + \rho \bm{\xi}_k\right) \right\rangle \\
        &\leq \left\langle \bm{\xi}_k + (\nabla_{\bm{v}}\bm{\xi}_k)^T(\bm{v}^* - \bm{v}_k), \pmb{\phi}_k + \rho\bm{\xi}_k \right\rangle
    \end{align*}
    where we use the convexity of $f_{\bm{y}}$ to produce the final inequality. Applying Lemma~\ref{lemma:full inner product bound prelim 1} produces the desired result.
\end{proof}

We now prove the desired bound on the augmented Lagrangian over one iteration of Algorithm~\ref{alg:CG-GAN}.

\begin{proof}[Proof of Proposition~\ref{prop:all variable update + full inner product bound}]
    Applying Corollary~\ref{corollary:dual-variable update bound} followed by Propositions~\ref{prop:w update bound},~\ref{prop:u update bound},~\ref{prop:z update bound}, and~\ref{prop:x update bound} and finally Lemma~\ref{lemma:bound ||zeta_k+1 - zeta_k||} gives
    \begin{align*}
        &\mathcal{L}_{k+1} = \overline{\mathcal{L}}_{k+1} + \sum_{j = 1}^4 R_j(\bm{\zeta}_j^{(k+1)}) \\
        &\s\s \leq \mathcal{L}_k + \theta\left\langle \bm{v}^*- \bm{v}_k, \nabla_{\bm{v}} \overline{\mathcal{L}}_k\right\rangle + \sum_{j = 1}^4 \left(\frac{\theta^2}{\alpha_j}\Delta_{j,k}^2 + \widehat{\gamma}_j\Delta_{j,k}^2\right) \\
        &\s\s + \theta\sum_{j = 1}^4 (R_j(\bm{\zeta}^*)-R_j(\bm{\zeta}_j^{(k)})).
    \end{align*}
    for $\widehat{\gamma}_1 = 2\tau_G^2(\rho^2+\sigma_0), \widehat{\gamma}_2 = 2[(\rho u_{\infty})^2 + (2\rho u_{\infty}z_\infty + 4\sigma_0 + \rho c_\infty)^2 + \sigma_0(1+2u_{\infty}^2)], \widehat{\gamma}_3 = 2z_{\infty}^2(\rho^2+2\sigma_0),$ and $\widehat{\gamma}_4 = 2\sigma_0$.

    Applying Lemma~\ref{lemma:full inner product bound} produces the desired result for $\beta_1 = \frac{L_G(4\sigma_0+\rho \xi_{1,\max})}{2}+\widehat{\gamma}_1$, $\beta_2 = \frac{4\sigma_0 + \rho\xi_{2,\max}}{2} + \widehat{\gamma}_2$, $\beta_3 = \frac{4\sigma_0 + \rho\xi_{2,\max}}{2} + \widehat{\gamma}_3$, and $\beta_4 = \widehat{\gamma}_4$ where $\xi_{j,\max} = \max_{k} \norm{\bm{\xi}_{j,k}}_2.$
\end{proof}

\subsection{Convergence (Proof of Theorem~\ref{thm:convergence})}\label{apndx:proof of thm:convergence}

We first establish a handful of preliminary results through the following lemmas in order to prove Theorem~\ref{thm:convergence}.

\begin{lemma}\label{lemma:theta choice prelim 3}
For $k\in\mathbb{N}$ assume that $\norm{\bm{\varphi}_{j,k} - \bm{\varphi}_j^*}_2^2 \geq \varphi_{j,\min}> 0$ and $\Delta_{3,k}^2 \geq p_3 > 0$ where $j\in\{1,2\}$. Let $\omega_1 > \frac{L_G\norm{\bm{\varphi}_1^*}_2}{\rho\nu_G^2}, \omega_2>\frac{\norm{\bm{\varphi}_2^*}_2}{\rho}, \omega_3 > \frac{\norm{\bm{\varphi}_2^*}_2}{\rho p_3},$ and $\omega_4 > 0$. Set $\omega_{5,1}\geq \frac{\omega_{1}\tau_G \Delta_{1,0}^2 + \omega_2\Delta_{2,0}^2}{\varphi_{1,\min}}$ and $\omega_{5,2} \geq \frac{\omega_3 \Delta_{(2,3),0}^2 + \omega_4\Delta_{4,0}^2}{\varphi_{2,\min}}$. Then
\begin{align*}
    \mathcal{L}_k - \mathcal{L}_* \geq \frac{1}{2}\sum_{i=1}^4 \kappa_i \Delta_{i,k}^2 - \frac{\mu}{\rho}
\end{align*}
for positive constants $\kappa_1 = \rho\omega_1\nu_G^2 -L_G\norm{\bm{\varphi}_1^*}_2, \kappa_2 = \rho\omega_2 - \norm{\bm{\varphi}_2^*}_2, \kappa_3 = \rho\omega_3p_3 - \norm{\bm{\varphi}_2^*}_2,$ $\kappa_4 = \rho\omega_4$, and $\mu = \left(1+\max\{\omega_{5,1}, \omega_{5,2}\}\right)\sum_{i = 1}^2 \left(\varphi_{i,\max}^2 + \norm{\varphi_i^*}_2^2\right)$.
\end{lemma}
\begin{proof}
Note that the first order optimality conditions of (\ref{eqn:iterative_opt}), with $F$ given in (\ref{eqn:iterative cost fnc generalized}), are $-\nabla_{\bm{v}} \overline{\mathcal{L}}_*\in \partial R_1 (\bm{x}^*)\times \partial R_2 (\bm{z}^*)\times \partial R_3 (\bm{u}^*)\times \partial R_4 (\bm{c}^*)$ and $\bm{\xi}^* = \bm{0}$ where $\partial R_i(\bm{\zeta}_i)$ is the subdifferential of $R_i$ at $\bm{\zeta}_i$. By the definition of a subdifferential
\begin{align*}
    \sum_{i=1}^4 \left[R_{\bm{\zeta}_i}(\bm{\zeta}_i^{(k)}) - R_{\bm{\zeta}_i}(\bm{\zeta}_i^*)\right] \geq \langle \bm{v}_k - \bm{v}^*, -\nabla_{\bm{v}} \overline{\mathcal{L}}_* \rangle.
\end{align*}

Thus, by (\ref{eqn:gradient of line L}), $\bm{\xi}^* = \bm{0}$, and convexity of $f_{\bm{y}}$, observe
    \begin{align}
        & \mathcal{L}_k - \mathcal{L}_* = \overline{\mathcal{L}}_k - \overline{\mathcal{L}}_* + \sum_{i=1}^4 \left[R_{\bm{\zeta}_i}(\bm{\zeta}_i^{(k)}) - R_{\bm{\zeta}_i}(\bm{\zeta}_i^*)\right] \nonumber \\
        &\resizebox{\columnwidth}{!}{${\displaystyle \geq  f_{\bm{y}}(\bm{c}_k) + \langle \pmb{\phi}_k, \bm{\xi}_k\rangle + \frac{\rho}{2}\norm{\bm{\xi}_k}_2^2 - f_{\bm{y}}(\bm{c}^*) - \langle \bm{v}_k - \bm{v}^*, \nabla_{\bm{v}} \overline{\mathcal{L}}_* \rangle}$} \nonumber \\
        &\geq  \langle \pmb{\phi}_k, \bm{\xi}_k\rangle + \frac{\rho}{2}\norm{\bm{\xi}_k}_2^2 + \langle (\nabla_{\bm{v}}\bm{\xi}^*)^T(\bm{v}^*-\bm{v}_k),  \pmb{\phi}^*\rangle \nonumber \\
        &\geq \langle \pmb{\phi}_k-\pmb{\phi}^*, \bm{\xi}_k\rangle + \frac{\rho}{2}\norm{\bm{\xi}_k}_2^2 \nonumber \\
        &\s\s -\frac{L_G\norm{\bm{\varphi}_{1}^*}_2}{2} \Delta_{1,k}^2 - \frac{\norm{\bm{\varphi}_2^*}_2}{2}\left(\Delta_{2,k}^2 + \Delta_{3,k}^2\right) \label{eqn:conv lemma 1 eqn 0}
    \end{align}
    where we applied Lemma~\ref{lemma:full inner product bound prelim 1} to furnish the final inequality.

    Next, by Assumption~\ref{assume:G} and $\bm{\xi}^* = \bm{0}$, observe
    \begin{align}
       & \norm{\bm{z}_k - G(\bm{x}_k) - \frac{\bm{\varphi}_1^*-\bm{\varphi}_{1,k}}{\rho}}_2^2 \nonumber \\
       &\s\s = \norm{(\bm{z}_k-\bm{z}^*) - \left(G(\bm{x}_k)-G(\bm{x}^*)\right) - \frac{\bm{\varphi}_1^*-\bm{\varphi}_{1,k}}{\rho}}_2^2 \nonumber \\
        &\s\s \geq \omega_2 \Delta_{2,k}^2 + \omega_1 \nu_G^2 \Delta_{1,k}^2 - \frac{\omega_{5,1}}{\rho^2} \norm{\bm{\varphi}_1^*-\bm{\varphi}_{1,k}}_2^2 \label{eqn:conv lemma 1 eqn 1}
    \end{align}
    and
    \begin{align}
        & \norm{\bm{c}_k - \bm{z}_k\odot\bm{u}_k - \frac{\bm{\varphi}_2^*-\bm{\varphi}_{2,k}}{\rho}}_2^2 \nonumber \\
        &\s\s = \norm{(\bm{c}_k-\bm{c}^*)- (\bm{z}_k\odot\bm{u}_k -\bm{z}^*\odot\bm{u}^*) - \frac{\bm{\varphi}_2^*-\bm{\varphi}_{2,k}}{\rho}}_2^2 \nonumber \\
        &\s\s \geq \omega_4\Delta_{4,k}^2 + \omega_{3}p_3\Delta_{3,k}^2 - \frac{\omega_{5,2}}{\rho^2}\norm{\bm{\varphi}_2^*-\bm{\varphi}_{2,k}}_2^2. \label{eqn:conv lemma 1 eqn 2}
    \end{align}
    Additionally, observe that
    \begin{align}
        \norm{\pmb{\phi}_k-\pmb{\phi}^*}_2^2 \leq 2\sum_{i = 1}^2 \left(\varphi_{i,\max}^2 + \norm{\varphi_i^*}_2^2\right). \label{eqn:conv lemma 1 eqn 3} 
    \end{align}

    Finally, combining
    \begin{align*}
        \resizebox{\columnwidth}{!}{${\displaystyle\langle \pmb{\phi}_k-\pmb{\phi}^*, \bm{\xi}_k\rangle + \frac{\rho}{2}\norm{\bm{\xi}_k}_2^2 = \frac{\rho}{2}\norm{\bm{\xi}_k - \frac{\pmb{\phi}^*-\pmb{\phi}_k}{\rho}}_2^2 - \frac{\norm{\pmb{\phi}_k-\pmb{\phi}^*}_2^2}{2\rho} }$} 
    \end{align*}
     with (\ref{eqn:conv lemma 1 eqn 0}), (\ref{eqn:conv lemma 1 eqn 1}), (\ref{eqn:conv lemma 1 eqn 2}), and (\ref{eqn:conv lemma 1 eqn 3}) produces the desired bound.    
\end{proof}

\begin{lemma}\label{lemma:theta choice prelim 5}
Let the assumptions of Lemma~\ref{lemma:theta choice prelim 3} hold and set $\{\kappa_j\}_{j=1}^4$ and $\mu$ as given in Lemma~\ref{lemma:theta choice prelim 3}.
Choose $\{(\alpha_j, \beta_j)\}_{j=1}^4$, from Proposition \ref{prop:all variable update + full inner product bound}, to satisfy $ \widehat{\beta} \leq \widetilde{\beta}$ where $\widehat{\beta} \coloneqq \sqrt{\max_{1\leq i\leq 4}\,\,\alpha_i\beta_i}$ and $\widetilde{\beta} \coloneqq \min_{1\leq i\leq 4}\,\, \frac{\alpha_i\kappa_i}{4}$. If $\sum_{j = 1}^4 \frac{\Delta_{j,k}^2}{\alpha_j} \geq \eta \geq \frac{\mu}{2\rho}(\widetilde{\beta} - \widehat{\beta})^{-1}$ then $\left(\mathcal{L}_k - \mathcal{L}_*\right)\left(2\sum_{j = 1}^4 \frac{\Delta_{j,k}^2}{\alpha_j}\right)^{-1} \geq \widehat{\beta}.$
\end{lemma}
\begin{proof}
   By Lemma~\ref{lemma:theta choice prelim 3}
   \begin{multline}
       \left(\mathcal{L}_k - \mathcal{L}_*\right)\left(2\sum_{j = 1}^4 \frac{\Delta_{j,k}^2}{\alpha_j}\right)^{-1}  \\
       \geq \left(\frac{1}{2}\sum_{j=1}^4 \kappa_j \Delta_{j,k}^2 - \frac{\mu}{\rho}\right)\left(2\sum_{j = 1}^4 \frac{\Delta_{j,k}^2}{\alpha_j}\right)^{-1} \label{eqn:bound lemma:theta choice prelim 5} \\
       \geq \widetilde{\beta} - \frac{\mu}{2\rho\eta} 
   \end{multline}
   applying $\eta \geq \frac{\mu}{2\rho}(\widetilde{\beta} - \widehat{\beta})^{-1}$ produces the desired bound.
\end{proof}

We now prove the main convergence result for our proposed A-CG-GAN algorithm
\begin{proof}[Proof of Theorem~\ref{thm:convergence}]
    Define
    \begin{align*}
        \theta_k = \min\left\{1, \sqrt{\left(\mathcal{L}_k - \mathcal{L}_*\right)^2\left(2\sum_{j = 1}^4 \frac{\Delta_{j,k}^2}{\alpha_j}\right)^{-2} - \widehat{\beta}^2}\right\},
    \end{align*}
    which is well-defined by Lemma~\ref{lemma:theta choice prelim 5} (where we take $\eta = \widehat{\eta}$) and satisfies $\theta_k\in [0,1]$ for every $k\in\mathbb{N}$.

    \textbf{Case 1:} $\theta_k = \sqrt{\left(\mathcal{L}_k - \mathcal{L}_*\right)^2\left(2\sum_{j = 1}^4 \frac{\Delta_{j,k}^2}{\alpha_j}\right)^{-2} - \widehat{\beta}^2}.$

    Using that $\sqrt{a-b}\geq \sqrt{a}-\sqrt{b}$ for all $a\geq b\geq 0$ observe
    \begin{align}
        &\theta_k^2\sum_{j = 1}^4 \frac{\Delta_{j,k}^2}{\alpha_j} + \theta_k (\mathcal{L}_* - \mathcal{L}) \nonumber \\
        &\leq \frac{\left(\mathcal{L}_k - \mathcal{L}_*\right)^2}{2}\left(2\sum_{j = 1}^4 \frac{\Delta_{j,k}^2}{\alpha_j}\right)^{-1} - \widehat{\beta}^2\sum_{j = 1}^4 \frac{\Delta_{j,k}^2}{\alpha_j} \nonumber \\
        &\s\s - \left(\mathcal{L}_k - \mathcal{L}_*\right)^2\left(2\sum_{j = 1}^4 \frac{\Delta_{j,k}^2}{\alpha_j}\right)^{-1} + \widehat{\beta}(\mathcal{L}_k - \mathcal{L}_*) \nonumber \\
        &\resizebox{\columnwidth}{!}{${\displaystyle\leq \widehat{\beta}(\mathcal{L}_k - \mathcal{L}_*) -\frac{\left(\mathcal{L}_k - \mathcal{L}_*\right)^2}{2}\left(2\sum_{j = 1}^4 \frac{\Delta_{j,k}^2}{\alpha_j}\right)^{-1} - \sum_{j = 1}^4 \beta_j \Delta_{j,k}^2 }$} \label{eqn:theta 1 bound}
    \end{align}
    where, in the final inequality, we used 
    \begin{align}
        \sum_{j = 1}^4 \beta_{j} \Delta_{j,k}^2 = \sum_{j = 1}^4 \alpha_j\beta_{j} \frac{\Delta_{j,k}^2}{\alpha_j} \leq \widehat{\beta}^2\sum_{j = 1}^4\frac{\Delta_{j,k}^2}{\alpha_j}. \label{eqn:theta choice prelim 7 eqn 1}
    \end{align}
    
    Now, by Proposition \ref{prop:all variable update + full inner product bound}, (\ref{eqn:theta 1 bound}), and (\ref{eqn:bound lemma:theta choice prelim 5}) observe
    \begin{align}
         & \mathcal{L}_{k+1} - \mathcal{L}_* \nonumber \\
         &\leq \mathcal{L}_k-\mathcal{L}_* + \sum_{j = 1}^4 \beta_{j} \Delta_{j,k}^2 + \theta_k^2\sum_{j = 1}^4 \frac{\Delta_{j,k}^2}{\alpha_j} + \theta_k (\mathcal{L}_* - \mathcal{L}) \nonumber \\
         &\leq (\mathcal{L}_k-\mathcal{L}_*)\left(1 -\frac{\mathcal{L}_k - \mathcal{L}_*}{2}\left(2\sum_{j = 1}^4 \frac{\Delta_{j,k}^2}{\alpha_j}\right)^{-1} + \widehat{\beta}\right) \nonumber \\
         &\leq \left(\mathcal{L}_k-\mathcal{L}_*\right)\left(1 - \frac{\widetilde{\beta}}{2} + \frac{\mu}{4\rho \widehat{\eta}} + \widehat{\beta}\right) \nonumber \\
         &\leq \left(\mathcal{L}_k-\mathcal{L}_*\right)\left(1-\widetilde{\beta}/8\right) \label{eqn:conv theta case 1 eqn 1}
    \end{align}
    where we use that $\widehat{\beta}\leq \frac{\widetilde{\beta}}{4}$ and $\widehat{\eta}\geq 2\mu (\rho \widetilde{\beta})^{-1}$ in the final inequality.

    \textbf{Case 2:} $\theta_k = 1.$

    By Proposition~\ref{prop:all variable update + full inner product bound} and (\ref{eqn:theta choice prelim 7 eqn 1}) observe
    \begin{align}
         & \mathcal{L}_{k+1} - \mathcal{L}_* \nonumber \\
         &\leq \mathcal{L}_k-\mathcal{L}_* + \sum_{j = 1}^4 \beta_{j} \Delta_{j,k}^2 + \sum_{j = 1}^4 \frac{\Delta_{j,k}^2}{\alpha_j} + (\mathcal{L}_* - \mathcal{L}) \nonumber \\
         &\leq (1+\widehat{\beta}^2)\sum_{j = 1}^4 \frac{\Delta_{j,k}^2}{\alpha_j} \nonumber \\
         &\leq \frac{\sqrt{1 + \widehat{\beta}^2}}{2}(\mathcal{L}_k - \mathcal{L}_*) \label{eqn:conv theta case 2 eqn 1}
    \end{align}
    where we use
    $1\leq \sqrt{\left(\mathcal{L}_k - \mathcal{L}_*\right)^2\left(2\sum_{j = 1}^4 \frac{\Delta_{j,k}^2}{\alpha_j}\right)^{-2} - \widehat{\beta}^2}$ in the final inequality. Note that $\sqrt{1+x^2} \leq 2-x$ for all $x\leq \frac{3}{4}$. Since $\widetilde{\beta}/4 \leq \frac{3}{4}$ then
\begin{align*}
    \sqrt{1+\widehat{\beta}^2} \leq \sqrt{1+\left(\widetilde{\beta}/4\right)^2} \leq 2- \widetilde{\beta}/4,
\end{align*}
which combined with (\ref{eqn:conv theta case 2 eqn 1}) gives (\ref{eqn:conv theta case 1 eqn 1}).

Hence, in both cases, $\mathcal{L}_{k+1}-\mathcal{L}_* \leq \left(1-\delta\right)\left(\mathcal{L}_k-\mathcal{L}_*\right)$. Applying this inequality recursively gives
\begin{align}
    \mathcal{L}_{k+1}-\mathcal{L}_* \leq \left(1-\delta\right)^{k+1}\left(\mathcal{L}_0-\mathcal{L}_*\right). \label{eqn:conv theta eqn 1}
\end{align}
Therefore, combining Lemma~\ref{lemma:theta choice prelim 3} and (\ref{eqn:conv theta eqn 1})
\begin{align*}
    \resizebox{\columnwidth}{!}{${\displaystyle \sum_{j = 1}^4 \frac{\Delta_{j,k}^2}{\alpha_j} \leq \frac{1}{4\widetilde{\beta}} \sum_{j = 1}^4 \kappa_j \Delta_{j,k}^2 \leq \frac{1}{4\widetilde{\beta}}\left(\left(1-\delta\right)^{k}\left(\mathcal{L}_0-\mathcal{L}_*\right) + \frac{\mu}{\rho}\right). }$}
\end{align*}
Finally, observing that $\widetilde{\beta}^{-1} \leq (4\widehat{\beta})^{-1}$ and $\frac{\mu}{4\rho\widetilde{\beta}} = \frac{\widehat{\eta}}{16}$ produces the desired result.
\end{proof}

\bibliographystyle{IEEEtran}
\bibliography{main}

\end{document}